\documentclass{emulateapj}
\usepackage{epsfig, natbib}
\usepackage{color}
\usepackage[colorlinks=true,citecolor=blue,linkcolor=blue]{hyperref}

\tightenlines

\countdef\decade=200
\advance\decade by \year
\countdef\hours=201	
\advance\hours by \time
\divide\hours by 60
\countdef\mins=202
\advance\mins by \hours
\multiply\mins by 60
\multiply\hours by 100
\countdef\miltime=203
\advance\miltime by \hours
\advance\miltime by \time
\advance\miltime by -\mins

\newcommand{\msun}{M_{\odot}}
\newcommand{\epsff}{\epsilon_{\rm ff}}
\newcommand{\tff}{t_{\rm ff}}
\newcommand{\tffgmc}{t_{\rm ff, GMC}}
\newcommand{\tfft}{t_{\rm ff, T}}
\newcommand{\epsdyn}{\epsilon_{\rm orb}}
\newcommand{\torb}{t_{\rm orb}}

\newcommand{\ltsim}{\protect\raisebox{-0.5ex}{$\:\stackrel{\textstyle <}
	{\sim}\:$}}

\begin{document}

\tabletypesize{\scriptsize}

\title{A Universal, Local Star Formation Law in Galactic Clouds, Nearby Galaxies, 
High-Redshift Disks, and Starbursts}

\shorttitle{Star Formation Laws}
\shortauthors{Krumholz \& Dekel}

\author{Mark R. Krumholz\altaffilmark{1},
Avishai Dekel\altaffilmark{2},
and
Christopher F. McKee\altaffilmark{3}
}
\altaffiltext{1}{Department of Astronomy, University of California, Santa Cruz, CA 95064; krumholz@ucolick.org}
\altaffiltext{2}{Racah Institute of Physics, The Hebrew University,
Jerusalem 91904, Israel; dekel@phys.huji.ac.il}
\altaffiltext{3}{Departments of Physics and Astronomy, University of California, Berkeley, CA 94720; cmckee@astro.berkeley.edu}

\begin{abstract}
Star formation laws are rules that relate the rate of star formation in a particular region, either an entire galaxy or some portion of it, to the properties of the gas, or other galactic properties, in that region. While observations of Local Group galaxies show a very simple, local star formation law in which the star formation rate per unit area in each patch of a galaxy scales linearly with the molecular gas surface density in that patch, recent observations of both Milky Way molecular clouds and high redshift galaxies apparently show a more complicated relationship, in which regions of equal molecular gas surface density can form stars at quite different rates. These data have been interpreted as implying either that different star formation laws may apply in different circumstances, that the star formation law is sensitive to large-scale galaxy properties rather than local properties, or that there are high density thresholds for star formation. Here we collate observations of the relationship between gas and star formation rate from resolved observations of Milky Way molecular clouds, from kpc-scale observations of Local Group galaxies, and from unresolved observations of both disk and starburst galaxies in the local universe and at high redshift. We show that all of these data are in fact consistent with a simple, {\it local, volumetric} star formation law. The apparent variations stem from the fact that the observed objects have a wide variety of 3D size scales and degrees of internal clumping,
so even at fixed gas column density the regions being observed can have wildly varying volume densities. 
We provide a simple theoretical framework to remove this projection effect, and we use it to show that
all the data, from small Solar neighborhood clouds with masses $\sim 10^3$ $\msun$ to sub-mm galaxies with masses $\sim 10^{11}$ $\msun$, fall on a single star formation law in which the star formation rate is simply $\sim 1\%$ of the molecular gas mass per local free-fall time. In contrast, proposed star formation laws in which the star formation timescale is set by the galactic rotation period are inconsistent with the data from the Milky Way and the Local Group, while those in which the star formation rate is linearly proportional to the gas mass above some density threshold fail both in the Local Group and for starburst galaxies.
\end{abstract}

\slugcomment{ApJ in press}


\section{Introduction}

The search for systematic relationships between the gas in galaxies and their star formation rates (SFRs) goes back to the seminal work of \citet{schmidt59a} and \citet{kennicutt89b}. Only in the last fifteen years, however, have observations advanced to the point where firm determinations of this relationship have become possible. \citet{kennicutt98a} showed that galaxies exhibit strong correlations between the surface density of star formation, the gas surface density, and the galactic rotation period. More recently, a number of authors have extended this analysis to $\sim$kpc-scale regions within several Local Group galaxies \citep{kennicutt07a, bigiel08a, leroy08a, blanc09a}. Their results have led to a picture of star formation in nearby galaxies with several important features. First, they find that star formation appears to be a function solely of {\it local} properties, with no evidence for systematic variations in the star formation law with respect to galactocentric radius, galactic rotation period, Toomre $Q$, or any other large-scale properties. Only the local surface density of molecular gas appears to matter. This is consistent with the complementary observation that the properties of star-forming molecular clouds in nearby galaxies show no systematic variation with galactic properties \citep{bolatto08a, fukui10a} (although there is some preliminary evidence that the GMC mass function might vary from galaxy to galaxy -- \citealt{rosolowsky05b, wong11a}). Models based on a local picture of star formation in molecular clouds have been reasonably successful at explaining, and in some cases predicting, these results \citep{krumholz05c, krumholz09b, ostriker10a}.

However, this picture has been complicated by more recent observations pushing to smaller scales and to higher redshifts. On small scales, \citet{evans09a}, \citet{lada10a}, and \citet{heiderman10a} show that molecular clouds within $\sim 1$ kpc of the Sun have SFR surface densities that are factors of $\sim 10$ greater than is found in (much larger) extragalactic regions of equal gas surface density. They propose that the star formation rate is determined by the mass in ``dense" gas, where dense can denote a threshold in either surface or volume density (see also \citealt{wu05a, wu10a}). In this model the local clouds show higher SFRs because they are above the threshold, while much of the gas seen in observations of nearby galaxies is below it.

In the distant universe, \citet{daddi10a} and \citet{genzel10a} compile samples of disk and starburst galaxies both locally and at high redshift, and show that there is a similar systematic offset: starburst galaxies typically have SFR surface densities that are a factor of $\sim 10$ higher than disk galaxies at equal gas surface density (however, as \citealt{ostriker11a} and \citealt{narayanan11b} point out, the disk starburst distinction is significantly enhanced by the use of single, different CO-H$_2$ conversion factors for discs and starbursts, which is almost certainly an oversimplification). While the distinction between disks and starbursts is not completely sharp (e.g.\ objects like M82 are ``weak" starbursts within disks), that two galaxies with the same gas surface density can display very different SFRs suggests that there must be some factor in addition to surface density that determines the SFR. Both \citet{daddi10a} and \citet{genzel10a} suggest that this factor has to do with some sort of dynamical time, and they argue in favor of it being the galactic orbital period. It is unclear by exactly what mechanism the orbital period affects the SFR.

A sensitivity to the galactic orbital period, or a density threshold, are difficult to reconcile with the star formation law observed in the Local Group data, which shows no evidence for either. It is not even clear how dependence on the orbital period would manifest on local scales, but, ultimately, the star formation law for galaxies as a whole must be the result of adding up numerous local patches. Furthermore, we note that numerical simulations of star formation are generally based on a purely local star formation law with no explicit dependence on galactic orbital period, and that simulations of entire galaxies never have the resolution to reach the proposed density thresholds of $\sim 10^4-10^5$ cm$^{-3}$ \citep[e.g.][to name a few]{springel03a, robertson08b, gnedin09a, bournaud10a, teyssier10a, ceverino10a, agertz11a, kuhlen11a}\footnote{In some of these simulations, e.g.\ \citet{bournaud10a} and \citet{hopkins11a}, gas does reach densities in this range, but these simulations also add an artificial pressure in high density gas to ensure that the Jeans length is well-resolved. This artificial pressure begins to dominate at densities above $n\approx 6 (\Delta x/100\mbox{ pc})^{-4/3}$ H cm$^{-3}$ \citep{teyssier10a}, where $\Delta x$ is the spatial resolution. For \citet{bournaud10a}, $\Delta x = 0.8$ pc, so artificial pressure dominates at densities above $3800$ H cm$^{-3}$. \citet{hopkins11a} use SPH simulations for which the resolution is spatially variable, but even for their highest resolution simulations $\Delta x \sim 2$ pc, corresponding to artificial pressure dominating at densities above $1100$ H cm$^{-3}$.
Thus, even if high density gas can appear in these simulations, its properties should be treated with great caution due to the effects of the artificial pressurization.}. Nonetheless, at least some of these simulations seem able to reproduce many of the observations on which the claims for a non-local star formation law are based \citep[e.g.][]{teyssier10a}. 

The goal of this paper is to alleviate this confusion by pointing out that all the data that have been thought to provide support for multiple star formation laws, sensitivity to the global orbital period, or density thresholds are in fact consistent with a single, simple volumetric star formation law with no thresholds and no direct dependence on the galactic orbital period. The apparent conflict between this model and the data stems from a failure to properly account for projection effects, a problem which has been noted before \citep{shetty08a}. We provide a simple method to account for these effects, which makes it possible to combine data across a wide range of size scales, from individual Milky Way clouds to entire starburst galaxies. The remainder of this paper is as follows. In Section \ref{sec:sflaws} we discuss three simple models of star formation, and develop observational predictions for each one. In Section \ref{sec:data} we compare these models to the available observational data. Finally, in Section \ref{sec:discussion} we discuss and summarize our results.

\section{Possible Star Formation Laws}
\label{sec:sflaws}

We consider three possible models for the star formation law: a local one in which the quantity that matters is the local volume density of gas, a global one in which star formation occurs on a timescale set by the galactic rotation period, and a third model in which the SFR is linearly proportional to the mass of gas above some density threshold. Our goal is to determine which, if any, of these proposed laws is capable of simultaneously explaining the Galactic, Local Group, and disk and starburst data at low and high redshift. For simplicity we limit our attention to regions where the gas is predominantly cold and molecular, and thus able to form stars. In low surface density or low metallicity regions where the gas is significantly atomic, thermal and chemical processes become dominant in determining where stars can form, and the gravitational potential of the stars and dark matter may have significant effects \citep{robertson08b, krumholz09b, krumholz11b, gnedin09a, gnedin10a, ostriker10a, krumholz11d, kim11a}, resulting in a much more complex star formation law.

\subsection{A Volumetric Star Formation Law}

\subsubsection{The Projected Star Formation Law}

A local volumetric star formation law is simply a function that maps a gas volume density $\rho$ to a volume density of star formation $\dot{\rho}_*$. One particularly simple hypothesis for this law is that the star formation rate is simply some fraction of the molecular gas mass per free-fall time, $\tff = \sqrt{3\pi/32 G \rho}$, so that
\begin{equation}
\label{eq:sflaw}
\dot{\rho}_* = f_{\rm H_2} \epsff \frac{\rho}{\tff} = f_{\rm H_2} \epsff \sqrt{\frac{32 G \rho^3}{3\pi}}
\end{equation}
where $f_{\rm H_2}$ is the fraction of the mass in molecular form\footnote{For simplicity throughout this paper we will adopt $f_{\rm H_2} = 1$, and where possible we will compare only to molecular gas masses. However, we retain the $f_{\rm H_2}$ factor in the equations to remind the reader that stars form only in molecular gas.} and $\epsff$ is a dimensionless measure of the star formation rate, and is constant or nearly so. \citet{krumholz05c} present a first-principles calculation that shows $\epsff \approx 0.01$ in any supersonically turbulent medium, with a very weak dependence on other quantities that we will ignore here for simplicity. \citet{padoan11a} argue for a slightly different functional dependence of $\epsff$ on the virial ratio and Mach number, but their overall values of $\epsff$ for the range of parameters relevant to real star-forming regions are only a factor of a few larger than the \citeauthor{krumholz05c} value. Any observational argument for an additional dependence of the star formation law on large-scale galactic quantities, or for density thresholds, must be able to invalidate the null hypothesis of a constant $\epsff$ in equation (\ref{eq:sflaw}). Note that there is some ambiguity in the choice of scale over which $\tff$ is to be measured. We adopt the \citet{krumholz05c} approach in which the relevant size scale is that corresponding to the outer scale of the turbulence that regulates the SFR. 

The difficulty in comparing a star formation law such as this to observations, particularly extragalactic ones, is that we generally do not have access to information about volume densities. Instead, we only have access to quantities measured in projection, and we can only evaluate the projected version of equation (\ref{eq:sflaw}),
\begin{eqnarray}
\dot{\Sigma}_* & = & f_{\rm H_2} \epsff \frac{\Sigma}{\tff}.
\label{eq:sflawproj}
\end{eqnarray}
It is important to note here that $\Sigma$ is the mean surface density of the region being observed,\footnote{Throughout this paper we adopt the convention that $\Sigma$ without subscripts refers to the surface density of whatever region is being observed, regardless of its scale. Values that are averaged over some particular physical scale independent of what is being observed will be subscripted.} whether it is a single giant molecular cloud (GMC) or an entire galaxy, but $\tff$ is the free-fall time evaluated at the density averaged over length scales comparable to the outer scale of the turbulence, regardless of the mean density of the region being observed. In a galaxy like the Milky Way with discrete molecular clouds, these two scales are the same only if the observation targets an individual cloud, which is $\sim 10-100$ pc in size in Milky Way-like galaxies. Almost no extragalactic observations reach this resolution. To give a concrete example of why this is significant, consider a simplified ISM similar to that of the Milky Way, but fully molecular in keeping with our approximations. In this ISM all the gas is in GMCs with surface and volume densities $\Sigma_{\rm GMC} \approx 100$ $\msun$ pc$^{-2}$ and $n_{\rm GMC} \approx 30$ cm$^{-3}$ \citep{mckee99a}. In contrast, when averaged over $\sim$kpc scales, the ISM in our example galaxy is similar to that near the Solar Circle, with a surface density $\Sigma_{\rm gal}\approx 10$ $\msun$ pc$^{-2}$ and a mean volume density $n_{\rm gal}\approx 1$ cm$^{-3}$ \citep{boulares90a}. The space between the molecular clouds is filled with much lower density gas that forms stars at a far lower rate\footnote{In the Milky Way the inter-cloud gas is atomic and thus does not form stars at all, but that does not matter for the purposes of this example.}, and thus contributes negligibly to the star formation rate of the galaxy. If we were to observe a $\sim$kpc-sized region of this ISM from outside the galaxy, the surface density entering equation (\ref{eq:sflawproj}) would be $\Sigma=\Sigma_{\rm gal}=10$ $\msun$ pc$^{-2}$, since this describes the amount of gas available to form stars. However, the density that determines the free-fall time $\tff$ is the GMC density $n_{\rm GMC} = 30$ cm$^{-3}$, not the mean ISM density $n_{\rm gal} = 1$ cm$^{-3}$; this corresponds to a factor of 5 difference in $\tff$. Similarly, if one observes a region smaller than a GMC with an even higher density, such as a protocluster gas clump, the free-fall time will be correspondingly shorter.

\subsubsection{Estimating the Free-Fall Time}
\label{sec:tff}

In order to evaluate the right hand side of equation (\ref{eq:sflawproj}), we must have a means of estimating $\tff$, or equivalently $\rho$, for the star-forming region from observables. One simple approach is to hypothesize that the relevant density is simply the galactic midplane density, and to estimate this by assuming that all galaxies have roughly the same scale height \citep[e.g.][]{elmegreen02a}. These two assumptions give $\dot{\Sigma}_* \propto \Sigma^{1.5}$. However, neither assumption is likely to be generally valid. As already noted, in the Milky Way the GMC density is $\sim 100$ times the mean midplane density. Nor are galactic scale heights uniform, as pointed out by \citet{shetty08a}, particularly if we are including sub-galactic regions in our sample. Instead, values range from $\sim 100$ pc in the Milky Way \citep{boulares90a} to $\sim 10$ pc in ULIRGs \citep{scoville97a, downes98a} to sizes as small as $\sim 1$ pc for individual molecular clouds. In the other direction, in high redshift disks the scale height is not directly measured, but is likely to be large since the gas velocity dispersion is $\sim 50$ km s$^{-1}$ \citep{cresci09a}, a factor of $\sim 6$ larger than the typical value in local disk galaxies. For such heterogenous samples, $\dot{\Sigma}_*$ will no longer be a single-valued function of $\Sigma$.

We must therefore turn to the problem of estimating the density and free-fall time of star-forming regions. For Galactic observations where individual GMCs can be resolved, the mean cloud density can be directly or nearly directly measured, and we defer further discussion of the Galactic case to Section \ref{sec:galdata}. For extragalactic observations that do not resolve GMCs, the problem is harder because we can only measure the surface density $\Sigma_{\rm gal}$ averaged over size scales of (at best) $\sim 1$ kpc to (at worst) the entire galactic disk. To handle this problem we follow the approach taken by \citet{krumholz09b} with slight modifications.

In nearby galaxies with low surface densities, star formation occurs in discrete, gravitationally-bound GMCs that are much denser than the mean of the surrounding ISM ($n\sim 100$ cm$^{-3}$ versus $n\sim 1$ cm$^{-3}$). These survive for $\sim 30$ Myr ($\sim 5-10$ free-fall times; \citealt{fukui09a}), and their properties are observed to be independent of the Galactic environment \citep{bolatto08a}. On the other hand, at the high surface densities found in starbursts, or even in normal disk galaxies at $z\sim 2$, the ISM is a continuous star-forming, turbulent medium. Some fraction of the mass is found in gravitationally-bound clumps, but these are only overdense by factors of $\sim 10-20$, and unlike Galactic GMCs there is no phase transition at their edges to decouple them from the turbulence in the ambient ISM \citep{dekel09a, ceverino10a, ceverino11a}. As a result the outer scale of the turbulence is the galactic scale height, and the relevant density is simply the midplane density, perhaps slightly enhanced due to clumping (c.f.~\citealt{ostriker11a}). (We will see later, however, that disk galaxies at $z\approx 2$ are only in this regime by a factor of a few.)

\subsubsection{The Giant Molecular Cloud Regime}

First consider the case for Milky Way-like galaxies, which we will refer to as the GMC case. Both observations \citep{solomon87a, bolatto08a} and theory \citep{krumholz06d, goldbaum11a} show that GMCs have mean surface densities that scatter with a factor of a few around $\sim 100$ $\msun$ pc$^{-2}$, independent of galactic environment or GMC mass. We adopt a fiducial value $\Sigma_{\rm GMC} = 85$ $\msun$ pc$^{-2}$, the mean found by \citet{bolatto08a}.\footnote{Note that most of the galaxies in the \citeauthor{bolatto08a} sample are dwarfs, and the spirals have slightly higher mean surface densities of $\sim 150$ $\msun$ pc$^{-2}$, similar to the value found by \citet{roman-duval10a} for the Milky Way. Most of the galaxies included in our sample here are not dwarfs, but we continue to use 85 $\msun$ pc$^{-2}$ for consistency with \citet{krumholz09b}.} There is a significant scatter about this value from one cloud to another \citep[e.g.][]{roman-duval10a}, as is expected from theoretical models \citep[e.g.][]{goldbaum11a}. However, the observations for which we need this estimate are necessarily averaging over a large number of clouds, and this averaging will reduce the scatter considerably -- for example, the galaxy-to-galaxy scatter in GMC surface density reported by \citet[their Table 4]{bolatto08a} is only $0.26$ dex.

The characteristic GMC mass is roughly the two-dimensional Jeans mass in the galactic disk \citep{kim02a, mckee07a},
\begin{equation}
M_{\rm GMC} = \frac{\sigma^4}{G^2 \Sigma_{\rm gal}},
\end{equation}
where $\sigma$ is the gas velocity dispersion and $\Sigma_{\rm gal}$ is the average surface density in the region of the galaxy where the GMCs form. For this GMC mass and surface density, the corresponding density and free-fall time are
\begin{eqnarray}
\rho_{\rm GMC} & = & \frac{3\sqrt{\pi}}{4} \frac{G \sqrt{\Sigma_{\rm GMC}^3 \Sigma_{\rm gal}}}{\sigma^2},
\nonumber \\
\tffgmc & = & \frac{\pi^{1/4}}{\sqrt{8}} \frac{\sigma}{G(\Sigma_{\rm GMC}^3 \Sigma_{\rm gal})^{1/4}}.
\label{eq:tffexgal1}
\end{eqnarray}
Note that we are simplifying somewhat by ignoring the possibility that GMCs could be compressed to somewhat higher densities by stellar gravity in regions where the stellar density is high \citep{ostriker10a}. This tends to occur only in dwarf galaxies or in the outer parts of spiral galaxies, which make little contribution to the total star formation rate budget.

In nearby disk galaxies $\sigma\approx 8$ km s$^{-1}$, with less than a factor of 2 variation either within a single galaxy or between different galaxies \citep{dib06a, walter08a, chung09a}, which means that $\tff$ is nearly independent of galactic properties in this regime. Of course $\sigma$ can be much larger in either starburst galaxies or in high redshift disk galaxies. For both disk galaxies at $z>0$ and starbursts at all redshift, we adopt $\sigma\approx 50$ km s$^{-1}$. However, this choice is largely irrelevant because most of these systems are in the second regime, to which we now turn.

\subsubsection{The Toomre Regime}

In galaxies with higher surface densities and star formation rates, molecular clouds cease to be very overdense and dynamically decoupled from the rest of the ISM. For example, the rotationally-supported giant clumps found in high-$z$ disks are overdense only by factors of $\sim 10$, and they contain only $\sim 20\%$ of the total molecular mass of the galaxy \citep{ceverino10a, ceverino11a}, compared with an overdensity of $\sim 100$ for $\sim 100\%$ of the molecular mass in Milky Way GMCs. In this case the mean density in the star-forming gas is set primarily by the weight of the ISM as a whole, rather than by the properties of dynamically-decoupled bound GMCs. A number of authors have proposed models to estimate the mean ISM density in this regime \citep{thompson05a, krumholz05c, krumholz09b, ostriker11a}, and we will adopt the estimate of \citet{krumholz05c}. The midplane pressure in a galactic disk of surface density $\Sigma_{\rm gal}$ is
\begin{equation}
P = \rho \sigma^2 = \phi_P \frac{\pi}{2} G \Sigma_{\rm gal}^2,
\end{equation}
where $\rho$ is the midplane density and the dimensionless factor $\phi_P$ is unity for a pure gas disk, and \citeauthor{krumholz05c} show $\phi_P \approx 3$ in real galactic disks that contain stars as well. Consequently,
\begin{equation}
\label{eq:rhoexgal2}
\rho \approx \frac{\pi \phi_P G \Sigma_{\rm gal}^2}{2\sigma^2}.
\end{equation}
The Toomre $Q$ for the gas is
\begin{equation}
\label{eq:q}
Q =  \frac{\sqrt{2(\beta+1)}\sigma\Omega}{\pi G \Sigma},
\end{equation}
where $\Omega = 2\pi/\torb$ is the angular velocity of galactic rotation, $\torb$ is the galactic orbital period, and $\beta = \partial \ln v_{\rm rot}/\partial \ln r$ is the logarithmic index of the rotation curve. A flat rotation curve corresponds to $\beta = 0$, while solid body rotation is $\beta=1$. If we now hypothesize that $Q\approx 1$, then combining equations (\ref{eq:rhoexgal2}) and (\ref{eq:q}) gives
\begin{equation}
\label{eq:tffexgal2}
\rho_{\rm T} = \frac{(\beta+1) \phi_P \Omega^2}{\pi G Q^2},
\quad
\tfft = \sqrt{\frac{3\pi^2 Q^2}{32(\beta+1)\phi_P}}\frac{1}{\Omega}.
\end{equation}
We refer to this as the Toomre case. Note that equation (\ref{eq:tffexgal2}) is the same estimate of $\tff$ as in \citet{krumholz05c}. 

The formation of giant clumps does not significantly alter these estimates. To demonstrate this, consider the example of a galaxy with a fraction $\epsilon_{\rm cl}$ of its ISM mass in giant clumps, which are overdense by a factor $f$ compared to the interclump medium. Typical values are $\epsilon_{\rm cl} = 0.2$ and $f=10$ \citep{ceverino10a, ceverino11a}. The mean ISM density is then $\overline{\rho} = \rho_{\rm cl} / (f-\epsilon_{\rm cl} f + \epsilon_{\rm cl})$, where $\rho_{\rm cl}$ is the density in the clumps. If each component obeys equation (\ref{eq:sflaw}), with a little algebra one can show that the total star formation rate in the galaxy is $\dot{M}_* = \epsilon_{\rm ff} M/t_{\rm ff}(\overline{\rho}) [\epsilon_{\rm cl}(f-\epsilon_{\rm cl}f+\epsilon_{\rm cl})^{1/2}+(1-\epsilon_{\rm cl})(1-\epsilon_{\rm cl}+\epsilon_{\rm cl}/f)^{1/2}]$, where $\tff(\overline{\rho})$ is the free-fall time evaluated at the mean density $\overline{\rho}$. Thus giant clumps enhance the star formation rate compared to a smooth ISM of the same mean density by the factor in square brackets. For the fiducial values $\epsilon_{\rm cl} = 0.2$ and $f=10$, this is only $1.3$.

To join the two regimes, we simply take the higher of the densities (and thus the smaller of the free-fall times) produced by equations (\ref{eq:tffexgal1}) and (\ref{eq:tffexgal2}):
\begin{equation}
\label{eq:tffexgal}
\tff = \min(\tffgmc, \tfft).
\end{equation}
This is equivalent to assuming that the density in the star-forming gas will be either the value produced by GMC self-gravity or the value produced by the pressure of the entire ISM, whichever is larger. This provides us with an estimate of $\tff$ in terms of extragalactic observables that we can use in equation (\ref{eq:sflawproj}).

\subsection{A Global Star Formation Law}

Alternately, one may posit that the star formation law in galaxies depends explicitly on global, large-scale galactic properties. The most common such star formation law is
\begin{equation}
\label{eq:globalsf}
\dot{\Sigma}_* = \epsdyn \frac{\Sigma}{\torb},
\end{equation}
where again $\Sigma$ is the mean surface density of the observed region, and $\epsdyn$ is the fraction of the mass converted to stars per galactic rotation period.\footnote{Note that this version of the star formation law is sometimes written using the crossing time $t_{\rm cross} = r/v_{\rm rot} = \torb/2\pi$ in the denominator in place of $\torb$, and that the phrase ``dynamical time" is sometimes used to mean both $\torb$ and $t_{\rm cross}$. To minimize confusion we will only use $\torb$ in this paper, and we adjust all published data to this convention.} Typical values based on observational fits are $\epsdyn \approx 0.1$. We do not include a factor $f_{\rm H_2}$ in equation (\ref{eq:globalsf}) as we do in equation (\ref{eq:sflaw}), because in these models it is generally assumed that star formation is regulated by global processes that do not care about the thermal or chemical state of the gas. Several theoretical models yield star formation laws of this kind, including those based on supernova regulation, cloud collisions, or large-scale gravitational instabilities \citep[e.g.][]{silk97a, tan00a, li05a, silk09a}. Note that this formulation is very similar to that of equation (\ref{eq:sflawproj}), except that the timescale on the right hand side is now the global dynamical time of the galaxy rather than a local free-fall time. For the purposes of comparing this star formation law to observations of entire galaxies, there is no need to worry about projection effects, since the local density and free-fall time no longer matter, and the surface densities on the two sides of equation (\ref{eq:globalsf}) can be integrated to their average values across the entire galaxy without changing the equation.

Since $\tff\sim \Omega^{-1}$ on galactic scales for high surface density galaxies (equation \ref{eq:tffexgal2}), this star formation law is identical to the volumetric one (equation \ref{eq:sflawproj}) for that case. This equivalence of the local and global star formation laws for disks as a whole has been noted by many authors \citep[e.g.][]{krumholz05c, leroy08a, genzel10a}, and so it is impossible to decide whether the more physically meaningful star formation law is local or global if the only information available is averages over entire galaxies at high surface density. However, the two models are very different when applied either to low surface density galaxies, or to the individual molecular clouds within them. In this case the local law predicts that the star formation timescale will then be set by the local free-fall time inside an individual GMC, while equation (\ref{eq:globalsf}) predicts that the star formation timescale will continue to depend on the galactic orbital period. It is not entirely clear in the latter case whether $\torb$ should be the orbital period computed at the outer edge of the star-forming disk of the galaxy or the local orbital period at a particular galactocentric radius. We explore both possibilities below.

\subsection{A Threshold Star Formation Law}

Yet a third proposed star formation law is one in which there is a volume or column density threshold for star formation \citep[e.g.][]{lada10a, heiderman10a}. Based on observations of the linear relation between mass of gas traced by HCN and star formation rate (\citealt{gao04a, wu05a,wu10a}; however, see \citealt{gao07a} and \citealt{krumholz07g} for evidence that the relation is not strictly linear), in these models it is generally assumed that the star formation rate scales linearly with the mass above the threshold. The origin of the threshold is not precisely specified, although one possible model for how it could arise comes from the photoionization-regulated star formation model of \citet{mckee89a}; however, it is not clear that this model is relevant in starburst galaxies where the UV photon mean free path is very small. Best fits for a threshold in local molecular clouds are generally at surface densities of $\sim 100$ $\msun$ pc$^{-2}$ \citep{lada10a}, which \citeauthor{lada10a} and \citeauthor{heiderman10a} argue is roughly equivalent to a volume density threshold of $n\sim 10^4 - 10^5$ cm$^{-3}$, although the conversion between volume and column density seems highly uncertain.\footnote{It is important to distinguish this very high threshold from the thresholds of $n\sim 1-10$ cm$^{-3}$ that are commonly used in numerical simulations of galaxies. The latter are a rough way of separating cold molecular gas from warm atomic gas that cannot form stars, and are not needed in simulations that actually model the atomic to molecular transition \citep{robertson08b, gnedin09a, kuhlen11a}. In contrast, the threshold proposed by \citet{heiderman10a} and \citet{lada10a} is one that would strongly suppress star formation even in $\sim 10$ K molecular gas.} Because the exact value and nature of the threshold is somewhat uncertain,
as is the star formation timescale in gas above the threshold, it is not entirely clear how to go about comparing these models to observations. However, as an example we test the model proposed by \citet{heiderman10a}, in which the star formation rate is given by
\begin{equation}
\label{eq:sfthresh}
\dot{\Sigma}_* = \frac{\Sigma_{\rm dense}}{t_{\rm dense}} = \frac{f_{\rm dense} \Sigma}{t_{\rm dense}},
\end{equation}
where ``dense" here refers to gas above the proposed threshold, $t_{\rm dense}$ is the constant star formation timescale in the dense gas, and $f_{\rm dense}$ is the fraction of gas above the density threshold. \citet{heiderman10a} do not give an explicit value for $t_{\rm dense}$, but their best-fit value for objects above their threshold (their Figure 10) corresponds to $t_{\rm dense} \approx 80$ Myr; this is similar to the value of $t_{\rm dense} = 83$ Myr obtained by \citet{wu05a, wu10a}, and so we adopt it. \citet{lada10a} finds a somewhat lower value $t_{\rm dense} \approx 20$ Myr. As we will see, this just provides an overall scaling that does not materially change the result. For Galactic molecular clouds the factor $f_{\rm dense}$ can be directly measured, and ranges from values $\ll 1$ for entire GMCs to $f_{\rm dense} = 1$ for objects selected based on high volume density or extinction thresholds. For extragalactic systems \citeauthor{heiderman10a}\ adopt
\begin{equation}
\label{eq:fdense}
f_{\rm dense} = 
\left\{
\begin{array}{ll}
0.078 (\Sigma_{\rm gal}/\Sigma_{\rm th})^{0.4}, \quad & 0.078 (\Sigma/\Sigma_{\rm th})^{0.4} < 1 \\
1, & {\rm otherwise}
\end{array}
\right.,
\end{equation}
where $\Sigma_{\rm th} = 129$ $\msun$ pc$^{-2}$ is the proposed threshold, and $\Sigma_{\rm gal}$ here is the mean surface density of the galaxy, averaged over $\sim$kpc scales.

\section{Comparison to Observations}
\label{sec:data}

We now compare the three proposed star formation laws to a data set consisting of Milky Way clouds in the Solar neighborhood, $\sim$kpc-scale regions in Local Group galaxies, and unresolved observations of disk galaxies and starbursts, both locally and at high redshift. We note that this data set represents one way of comparing proposed star formation laws to observations. Another approach, which we discuss below in Section \ref{sec:molcorrelations}, is to correlate the star formation rate with luminosities of various molecular lines; we do not discuss this approach in detail here, since it has been the subject of numerous earlier papers.

A second note that applies to all the extragalactic data involves the CO $\alpha$-factor, the conversion between the measured CO luminosity and the H$_2$ mass. There is both strong observational \citep{solomon97a, downes98a, tacconi08a} and theoretical \citep{narayanan11a, narayanan11b} evidence that this factor is smaller in starbursts and sub-mm galaxies than it is in star-forming disks. We follow \citet{daddi10a} in using $\alpha_{\rm CO} = 0.8$ $\msun/(\mbox{K km s}^{-1}\mbox{ pc}^2)$ in starbursts at all redshifts, $\alpha_{\rm CO} = 4.6$ $\msun/(\mbox{K km s}^{-1}\mbox{ pc}^2)$ in disk galaxies at $z=0$, and $\alpha_{\rm CO} = 3.6$ $\msun/(\mbox{K km s}^{-1}\mbox{ pc}^2)$ in disk galaxies at high redshift, and we adjust all data to these choices. We have verified that, if we instead adopt the values favored by \citet{genzel10a} ($\alpha_{\rm CO} = 1.0$ $\msun/(\mbox{K km s}^{-1}\mbox{ pc}^2)$ in starbursts and $\alpha_{\rm CO} = 3.2$ $\msun/(\mbox{K km s}^{-1}\mbox{ pc}^2)$ in all disks regardless of redshift), the results do not change significantly. In reality, strictly bimodal values of $\alpha_{\rm CO}$ for disks versus starbursts are a crude approximation, and instead $\alpha_{\rm CO}$ should vary continuously as a function of galaxy properties \citep{narayanan11b}.

\subsection{Galactic Molecular Clouds}
\label{sec:galdata}

\begin{figure*}
\epsscale{1.18}
\plotone{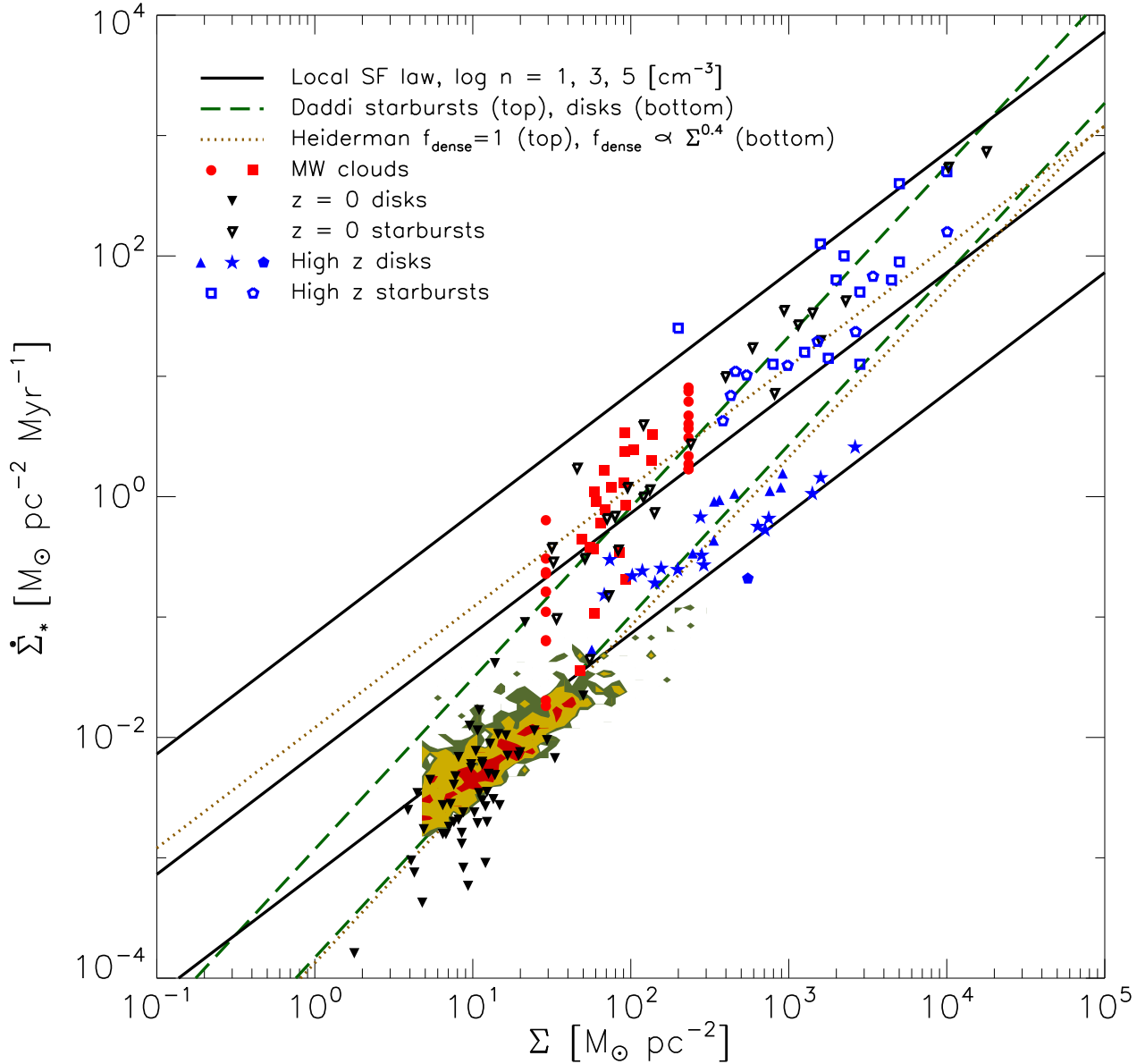}
\caption{
\label{fig:sigmasigma}
\scriptsize
Star formation surface density $\dot{\Sigma}_*$ versus observed gas surface density $\Sigma$. Contours and symbols show data, following the scheme that contours represent resolved observations of Local Group galaxies, filled symbols represent Galactic clouds (red) or disk galaxies at $z=0$ (black) or high-$z$ (blue), and open symbols represent starbursts at $z=0$ (black) or high-$z$ (blue). Solid lines show theoretical models: black lines show the volumetric star formation law, equation (\ref{eq:sflawproj}), evaluated using $\epsff = 0.01$ and 
volume densities $n = 10^1, 10^3$, and $10^5$ cm$^{-3}$ (bottom to top). Green dashed lines show the sequence of disks (lower) and sequence of starbursts (upper) models from by \citet{daddi10a}; these have slopes of 1.42. Brown dotted lines show the threshold model of \citet{heiderman10a}, evaluated using $f_{\rm dense} = 1$ (upper line) and $f_{\rm dense} \propto \Sigma^{0.4}$ (lower line; equation \ref{eq:fdense}), as indicated. The individual data sources are: Solar neighborhood molecular clouds from \citet[red squares]{heiderman10a} and \citet[red circles, at high and low $\Sigma$ corresponding to clouds defined by $A_K = 0.1$ mag and $A_K=0.8$ mag contours]{lada10a}; individual kpc-sized regions from in Local Group galaxies from THINGS \citep[contours, with the contour levels representing 1, 2, 5, and 10 data points from highest to lowest]{bigiel08a}; $z=0$ disk galaxies \citep[black filled downward triangles]{kennicutt98a}, $z=0$ starbursts \citep[black open downward triangles]{kennicutt98a}, $z=1-2.3$ BzK-selected galaxies \citep[blue filled stars]{tacconi10a}, $z=0.5$ and $z=1.5$ BzK-selected galaxies \citep[blue filled triangles]{daddi08a, daddi10b}, $z=1.2 - 2.3$ star-forming galaxies \citep[blue filled pentagons]{genzel10a}, and $z=1 - 3.5$ sub-mm galaxies (\citealt{bouche07a}, blue open squares; \citealt{genzel10a}, blue open pentagons). Note that the \citet{lada10a} clouds all line up at the same $\Sigma$ because they are defined by a column density threshold.
}
\end{figure*}

For Galactic clouds, the quantities appearing on the right hand side of equation (\ref{eq:sflawproj}) that can be measured most directly from dust absorption are the gas mass $M$ and projected area $A$, which can be combined to yield a surface density. The line of sight depth is difficult to determine for individual clouds, so we cannot directly measure the volume density. However, if we make the simplest possible assumption that the clouds are spherical, then the mean density is $\rho = (3\sqrt{\pi}/4) M/A^{3/2}$. In reality this assumption probably leads us to systematically underestimate the density in large clouds. These tend to be filamentary rather than spherical, and given random orientations a filamentary object is likely to have a line of sight depth smaller than $2\sqrt{A/\pi}$, the value implicitly assumed in our spherical assumption. However, since correcting for this effect would require us to know the intrinsic aspect ratios of molecular clouds, we retain the spherical assumption for simplicity and uniformity, and note below where it likely produces error.

The remaining quantity that appears on the left hand side of equation (\ref{eq:sflawproj}) is the SFR. This can be measured, but only in regions that are sufficiently old. Estimates of the SFR in Galactic sources are generally based on either number counts of young stellar objects \citep[YSOs;][]{evans09a, heiderman10a, lada10a} or measurements of the infrared or radio luminosity \citep{mooney88a, wu05a, wu10a, murray10a}. All of these methods rely on the assumption that the population is in statistical equilibrium between new objects forming and old ones disappearing, e.g.\ between massive stars forming and massive stars leaving the main sequence \citep{krumholz07e}. This requires that the region being observed have an age spread larger than the lifetimes of the objects in question -- roughly 2 Myr for class II YSOs, and $\sim 4$ Myr for the massive stars that dominate radio and infrared luminosities. While this is almost always true for extragalactic observations, it may not be for Galactic observations that target much smaller objects with shorter dynamical times. For this reason, we do not consider published estimates of the SFR for objects with dynamical times $\ll 1$ Myr (e.g.~the compact regions surveyed by \citealt{wu10a} or the high density sample of \citealt{heiderman10a}, which have crossing times $\sim 0.1$ Myr). These estimates are almost certainly unreliable.\footnote{One could avoid this problem by making an independent measurement of the age of the stellar population, e.g.\ by placing stars on the HR diagram and using pre-main sequence evolutionary tracks. An age plus either a number count or an IR or radio luminosity yields a unique SFR even for young ages. However, age data are not available for the vast majority of the objects in the high density samples of \citealt{wu10a} and \citet{heiderman10a}.}

Given these constraints, we take our Galactic data from two samples of nearby molecular clouds: those of \citet{heiderman10a} and \citet{lada10a}. \citet{heiderman10a} report cloud masses, star formation rates, and areas, and we use these to estimate $\Sigma$, $\dot{\Sigma}_*$, $\rho$, and $\tff$ as described above.

\citet{lada10a} report the gas mass and the total star formation rate based on number of YSOs enclosed within a contour of $K$-band extinction $A_K = 0.1$ mag, corresponding to a visual extinction $A_V = 0.89$ mag (using the same extinction law adopted by \citeauthor{lada10a}) and a surface density of $14.5$ $\msun$ pc$^{-2}$. The mean surface density of material within this contour is a factor of 2 larger than this \citep[also C.~Lada, 2011, priv.~comm.]{lombardi10a}. We therefore compute the corresponding area using $A = M/\Sigma$ with $\Sigma=29$ $\msun$ pc$^{-2}$, and use this value to compute $\dot{\Sigma}_*$, $\rho$, and $\tff$. \citet{lada10a} also report the gas mass within a contour $A_K = 0.8$ mag, corresponding to a visual extinction $A_V = 7.1$ mag, and a mean surface density $\Sigma \approx 230$ $\msun$ pc$^{-2}$ (again assuming that the mean is twice the threshold value). Unfortunately the full data set does not contain sufficient positional information on the YSOs to determine the number within the $A_K = 0.8$ mag contour, but in a subset of the data for which positional information is available, roughly $1/4-1/2$ of those within the $A_K = 0.1$ mag contour also lie within the $A_K = 0.8$ mag contour (C.~Lada, 2011, priv.~comm.). We therefore estimate the SFR within the $A_K = 0.8$ mag contour by assuming its value to be $1/3$ that of the full cloud.

Comparison of this data to the global star formation law, equation (\ref{eq:globalsf}), requires that we estimate $\torb$. As noted above, we consider two possibilities. One is $\torb$ evaluated at the galactocentric radius of the molecular clouds, which is roughly equal to the $r \approx 8$ kpc radius of the Sun  \citep{ghez08a, gillessen09a} since most of the clouds are closer than 1 kpc. We adopt a flat rotation curve at $v_{\rm rot} = 220$ km s$^{-1}$ \citep{fich89a}, which then gives $\torb = 2\pi r/v_{\rm rot} = 220$ Myr. Alternately, we can use the radius at the edge of the star-forming disk. This is somewhat ill-defined, but to maximize the difference from our previous $\torb$ value, we adopt the large value $r=15$ kpc, giving $\torb = 420$ Myr.

We summarize all the data, and the values of $\rho$ and $\tff$ that we derive from it, in the Appendix, Table \ref{tab:galdata}. It is worth noting that the values of $\rho$ in several of the \citet{lada10a} $A_K = 1$ clouds are almost certainly too small as a result of the systematic error described above. For example, our spherical assumption gives Orion A a line of sight depth of 50 pc, while the true value is probably a factor of 10 lower. We note in Table \ref{tab:galdata} values that are likely to be discrepant.

\subsection{Resolved Observations of Local Group Galaxies}

Our data set for resolved observations of Local Group galaxies is taken from The H~\textsc{i} Nearby Galaxy Survey (THINGS; \citealt{walter08a, bigiel08a, leroy08a}), supplemented with CO measurements from the HERA CO Line Extrgalactic Survey (HERACLES; \citealt{leroy09a}). THINGS plus HERACLES provide measurements of the surface densities of total gas, atomic gas, molecular gas, and star formation rate in $\sim$kpc-sized regions over a number of nearby galaxies. Since we are not treating the regime where the star formation law is set predominantly by the atomic to molecular transition (see Section \ref{sec:sflaws}), we use $\Sigma_{\rm gal} = \Sigma_{\rm H_2}$. Unfortunately we do not have access to pixel-by-pixel values of angular velocity $\Omega$; for these we only have azimuthal averages, and corresponding azimuthal averages of $\Sigma_{\rm H_2}$ and $\dot{\Sigma}_*$. We therefore use the pixel-by-pixel values when comparing $\dot{\Sigma}_*$ and $\Sigma_{\rm H_2}$, and the azimuthally-averaged values when comparing $\dot{\Sigma}_*$ and $\Sigma_{\rm H_2}/\torb$. To estimate $\tff$, we use the pixel data and compute the local free-fall time using equation (\ref{eq:tffexgal}), since, as we will see in Section \ref{sec:multiple}, local non-starburst galaxies are in the regime where $\tffgmc \la \tfft$.

\subsection{Unresolved Observations}

Our data set for unresolved observations (those in which only a single value is assigned to the entire galaxy) consists of normal disk galaxies and starbursts in the local universe taken from \citet{kennicutt98a}, and a collection of high-redshift systems compiled by \citet{daddi10a} and \citet{genzel10a}. These include star-forming disk galaxies at $z\approx 0.5-2.3$ \citep{daddi08a, daddi10b, tacconi10a, genzel10a}, which constitute the bulk of the star-forming galaxies at $z\sim 2$ \citep{dekel09b}, and sub-mm galaxies from $z=1-3.5$ \citep{bouche07a, genzel10a}.\footnote{Where a given object appears multiple times in the above references, we plot it only once.} All these samples include measurements of $\Sigma$ and $\Omega$. For all objects we estimate the local free-fall time using equation (\ref{eq:tffexgal}). In evaluating these equations we adopt $Q=1$ and $\beta = 0$ for disk galaxies, and $Q=1$ and $\beta = 1$ for starbursts and sub-mm galaxies, relying on the classifications provided by \citet{daddi10a} and \citet{genzel10a} to determine whether a particular galaxy is a disk or starburst, though, as noted above, the distinction is not sharp. This is equivalent to assuming a flat rotation curve for the large disk galaxies, and a solid-body rotation curve for the more compact starbursts, which are generally within the solid body rotation region of their galaxies. We note that $Q$ can be driven somewhat below unity by accretion from the intergalactic medium or by mergers \citep[e.g.][]{kim11a}. However, none of these choices has a large effect, and adopting different values of $\beta$ between 0 and 1, or $Q$ within a factor of a few of unity, yields qualitatively identical results.

We summarize the $z=0$ and $z>0$ unresolved galaxy data sets in the Appendix, Tables \ref{tab:unresolved_local} and \ref{tab:unresolved_hi_z}.

\subsection{Comparing Observations to Models}

\begin{figure*}
\epsscale{1.18}
\plotone{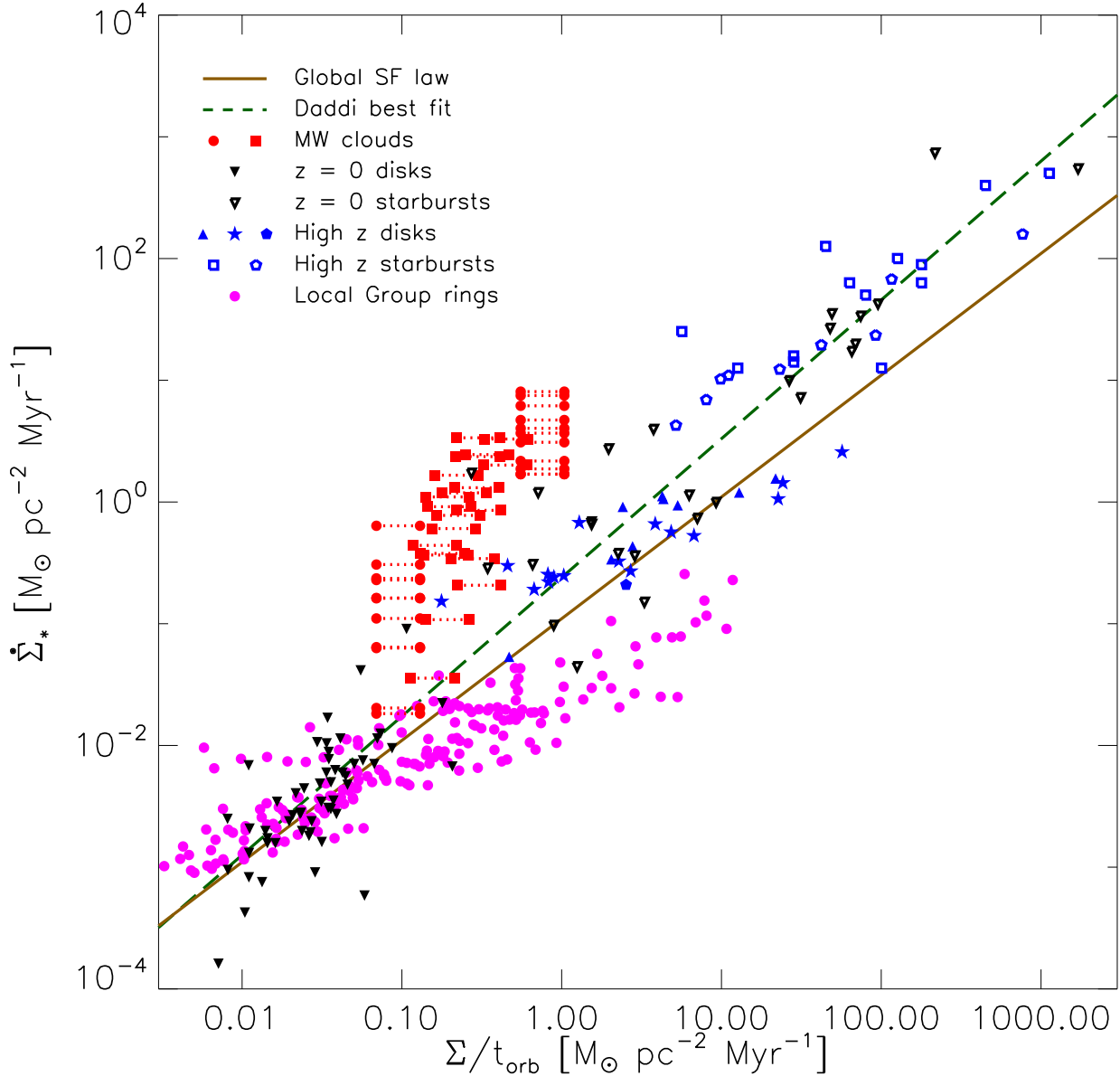}
\caption{
\label{fig:sigmatdyn}
Star formation surface density $\dot{\Sigma}_*$ versus gas surface density over galactic orbital period $\Sigma/\torb$. All symbols are as in Figure \ref{fig:sigmasigma}, except for the addition of the azimuthally-averaged rings in Local Group galaxies from THINGS/HERACLES \citep[magenta filled circles]{leroy08a, leroy09a}. The dynamic range on both axes is the same as in Figure \ref{fig:sigmasigma} in order to facilitate comparison. The red symbols connected by dotted lines represent the same Galactic molecular clouds, with the lower $\Sigma/\torb$ corresponding to $\torb$ evaluated at $r=15$ kpc, and the higher evaluated at $r=8$ kpc. The solid brown line is the best fit given by \citet{kennicutt98a}, which corresponds to equation (\ref{eq:globalsf}) with $\epsdyn = 0.11$. The dashed green line is the best fit of \citet{daddi10a} to the extragalactic data, which has a slope of $1.14$. Note that all the Milky Way data lies well above the fit line, while much of the Local Group data lies well below it.
}
\end{figure*}

\begin{figure*}
\epsscale{1.18}
\plotone{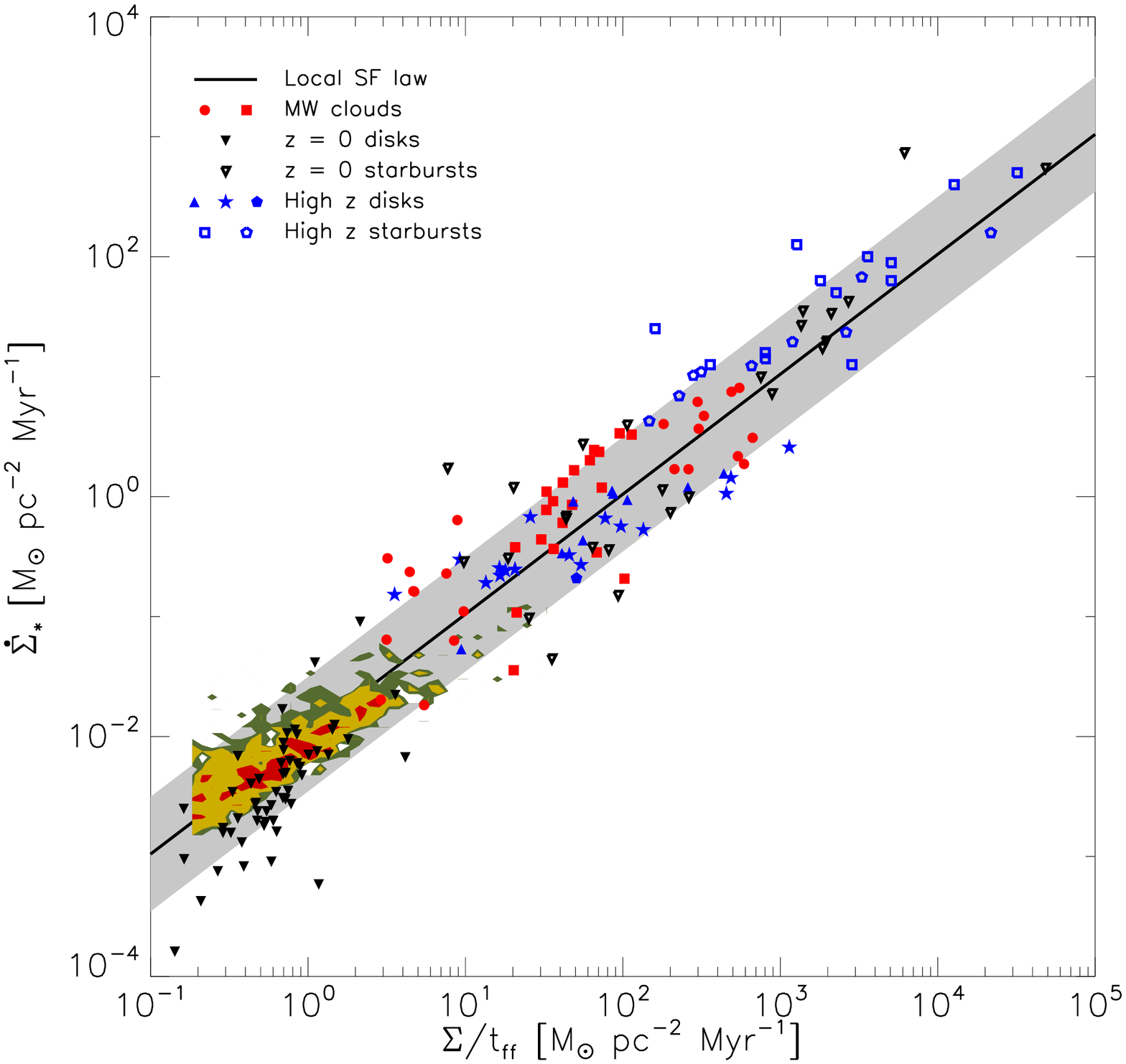}
\caption{
\label{fig:sigmal}
Star formation surface density $\dot{\Sigma}_*$ versus $\Sigma/t_{\rm ff}$. All symbols are as in Figure \ref{fig:sigmasigma}, and the dynamic range on both axes is the same in order to facilitate comparison. The solid line represents the local volumetric star formation law, equation (\ref{eq:sflawproj}), evaluated with the best-fit value $\epsff = 0.01$, and the gray band shows a factor of 3 range about this. Free-fall times for all objects are estimated as described in Section \ref{sec:tff}. All the data are consistent with a universal star-formation law, including the star forming regions in the Milky Way and the Local Group. Note that the Galactic clouds that lie above the fit at low $\Sigma/\tff$ are those most likely to be affected by the geometric errors described in Section \ref{sec:galdata}, so these points should be treated as uncertain.
}
\end{figure*}

In Figure \ref{fig:sigmasigma} we plot $\dot{\Sigma}_*$ versus $\Sigma$ for all of the Galactic and extragalactic data. We also overplot the best fits obtained by \citet{daddi10a} for disks and starbursts, the threshold model of \citet{heiderman10a}, and the projected volumetric star formation law, equation (\ref{eq:sflawproj}), evaluated with $\epsff = 0.01$ and 
volume densities $n=10^1$, $10^3$, and $10^5$ cm$^{-3}$. If we consider only the extragalactic data, we see that the observations appear to fall onto two separate sequences, one describing disk galaxies and one describing starbursts, as proposed by \citet{daddi10a} and \citet{genzel10a}. 
However, the Galactic observations do not follow this pattern. Instead, they lie systematically above even the starburst fit to the extragalactic sources, despite the fact that the sample consists of molecular clouds from the Milky Way, a disk galaxy, and that most of the clouds included are not regions of particularly vigorous star formation. Instead, they are predominantly small, weakly star-forming clouds like Taurus, Perseus, and Chameleon. Thus the sequence of disks and sequence of starbursts fits work well for the extragalactic data, but fail for the Galactic data.

The problem is reversed for the threshold model. If we evaluate this model with $f_{\rm dense} \propto \Sigma^{0.4}$, as proposed by \citet{heiderman10a} for extragalactic observations, the curve is nearly identical to the \citet{daddi10a} sequence of disks, except at very high surface densities, where it flattens. This model clearly fails for the starbursts. If we instead adopt $f_{\rm dense} = 1$, then the curve passes through the center of the \citeauthor{heiderman10a}\ Galactic sample, but is a poor match to all of the extragalactic data. Indeed, it is important to note that there is no single-valued function $f_{\rm dense}(\Sigma)$ that will make the threshold model agree with the observations. Of course one could propose a more complicated functional form for $f_{\rm dense}$ to force agreement, but in the absence of a theoretical model capable of explaining why $f_{\rm dense}$ should vary in this way, such a function would have no predictive power. Moreover, we note that, even with $f_{\rm dense} = 1$, the threshold model substantially underpredicts the SFR in most of the starbursts. One could attempt to remedy this by making $t_{\rm dense}$ small enough so that the $f_{\rm dense} = 1$ line would be safely above even the brightest starbursts. However, in this case $t_{\rm dense}$ would be so small than even the \citet{lada10a} $A_K = 0.8$ mag data would lie below the $f_{\rm dense} = 1$ line; there is no single value of $t_{\rm dense}$ that can simultaneously match this data set and the brightest sub-mm galaxies. Thus in order to fit both these data sets the dense gas depletion time $t_{\rm dense}$ would have to change. Since the constancy of $t_{\rm dense}$ is the basis of the entire model, this failure would appear to definitively rule out the model in its current form.

The model could potentially be saved by replacing $t_{\rm dense}$ with $100 t_{\rm ff}$, where $t_{\rm ff}$ is the free-fall time evaluated at the mean density of gas above the purported threshold. This would allow the star formation rate to increase in the densest starbursts, as it does in the local volumetric law, and as appears to be required by the data. However, even with this alteration there remains the problem that there is no independent way to predict $f_{\rm dense}$ from observables. Thus this model has extremely limited predictive power.

Figure \ref{fig:sigmatdyn} shows the star formation rate as a function of $\Sigma/\torb$, together with the best fit relation of \citet{daddi10a} and the global star formation law, equation (\ref{eq:globalsf}), evaluated with $\epsdyn=0.11$, the best-fit value from \citet{kennicutt98a}. We see that, while the global star formation law provides a reasonable fit to the unresolved extragalactic data, and the \citet{daddi10a} fit (which, unlike equation (\ref{eq:globalsf}) allows the slope to vary arbitrarily) agrees with the data even better, neither agrees at all with either the Galactic or resolved Local Group data. Instead, the Galactic data lie systematically above the extragalactic relation, the Local Group data lie mostly below it, and the indicated slopes for all three data sets are different. Moreover, we note that Figure 17 of \citet{leroy08a} shows that, in the resolved Local Group data, the ratio $\dot{\Sigma}_*/\Sigma_{\rm H_2}$ in fact remains invariant as $\torb$ changes by almost an order of magnitude. This implies that the slight positive slope displayed by the Local Group data in Figure \ref{fig:sigmatdyn} arises just because $\dot{\Sigma}_*$ and $\Sigma_{\rm H_2}$ are well-correlated, and dividing by an additional factor of $\torb$ on the $x$ axis does not completely destroy that correlation. In any event, it is evident that a star formation law that depends on the global galactic rotation period, such as equation (\ref{eq:globalsf}) does not provide a good description of star formation in molecular clouds in the Milky Way, or in $\sim$kpc-sized regions in Local Group galaxies.

Finally, in Figure \ref{fig:sigmal} we plot $\dot{\Sigma}_*$ versus $\Sigma/\tff$, the quantity that is expected to control the star formation rate for a local, volumetric star formation law. It is immediately apparent that this relation provides a far better fit than either of the alternatives. The Galactic and extragalactic data now all lie on the same relation. As we have already noted, our estimate of $\tff$ makes it proportional to $\torb$ for galaxies whose high surface densities put them in the Toomre regime, so with whole-galaxy data alone it is difficult to distinguish between the local and global star formation laws, equations (\ref{eq:sflawproj}) and (\ref{eq:globalsf}). However, the addition of the Galactic and Local Group data clearly breaks this degeneracy in favor of the local star formation law.


\begin{deluxetable}{lccc}
\tablecaption{Best Fit Parameters\label{tab:fits}}
\tablehead{
\colhead{Data Included in Fit} &
\colhead{$\eta$} &
\colhead{$q$} &
\colhead{Scatter\tablenotemark{a}} \\
}
\startdata
\cutinhead{Fits to figure \ref{fig:sigmasigma}, functional form\tablenotemark{b} $\dot{\Sigma}_* = \eta \Sigma^q$} \\
Unresolved extragalactic disks & $0.00019$ & $1.31$ & $2.2$ \\
Unresolved extragalactic starbursts & $0.0027$ & $1.26$ & $2.3$ \\
All data & $0.016$ & $0.73$ & $20$ \\
\cutinhead{Fits to figure \ref{fig:sigmatdyn}, functional form\tablenotemark{b} $\dot{\Sigma}_* = \eta (\Sigma/\torb)^q$} \\
All unresolved extragalactic & 0.23 & 1.13 & 2.7 \\
All unresolved extragalactic, $q=1$ & 0.22 & 1.0 & 3.0 \\
All & 0.50 & 0.48 & 21 \\
\cutinhead{Fits to figure \ref{fig:sigmal}, functional form $\dot{\Sigma}_* = \eta (\Sigma/\tff)$} \\
All & 0.01 & \nodata & 2.8
\enddata
\tablenotetext{a}{The scatter given is a multiplicative factor, so a scatter of unity indicates perfect agreement between data and fit.}
\tablenotetext{b}{In these fits $\dot{\Sigma}_*$ has units of $\msun$ pc$^{-2}$ Myr$^{-1}$, $\Sigma$ has units of $\msun$ pc$^{-2}$, and $\torb$ has units of Myr.}
\end{deluxetable}


We can demonstrate the superiority of the volumetric star formation law quantitatively by fitting to the data shown in Figures \ref{fig:sigmasigma} -- \ref{fig:sigmal}. We summarize the fit parameters in Table \ref{tab:fits}. For Figure \ref{fig:sigmasigma}, if we fit a powerlaw function of the form
\begin{equation}
\dot{\Sigma}_*\;[\msun\mbox{ pc}^{-2}\mbox{ Myr}^{-1}] = \eta (\Sigma\;[\msun\mbox{ pc}^{-2}])^{q}
\end{equation}
to the unresolved extragalactic disks and starbursts separately,\footnote{For this and the other fits we discuss below, we do not include the THINGS data, because it is not clear how to weight them together with the observations of single objects.} the best fitting slopes are $q=1.31$ and $1.26$ for the disks and starbursts, respectively. The scatter in these fits is modest, a factor of $2.2$ and $2.3$. The corresponding best-fit parameters and scatter obtained by \citet{daddi10a} for their disk and starburst data are quite similar. However, if we attempt to fit all the data simultaneously, Galactic and extragalactic, the fit is far different and far worse: slope $q=0.73$, factor of $20$ scatter. The failure of a powerlaw fit between $\dot{\Sigma}_*$ and $\Sigma$ for the extragalactic data including both disks and starbursts is consistent with the findings of \citet{daddi10a} and \citet{genzel10a}, and here we see that the inclusion of the Galactic data further compounds the problem.

For Figure \ref{fig:sigmatdyn}, if we limit the fit to the unresolved extragalactic data, and fit a powerlaw of the form
\begin{equation}
\label{eq:tdynfit}
\dot{\Sigma}_*\;[\msun\mbox{ pc}^{-2}\mbox{ Myr}^{-1}] = \eta \left(\frac{\Sigma}{\torb}\;[\msun\mbox{ pc}^{-2}\mbox{ Myr}^{-1}]\right)^{q},
\end{equation}
the best fit slope is $q=1.13$, with a factor of 2.7 scatter. Again, these values are nearly identical to those obtained by \citet{daddi10a}. If we fix the slope to $q=1.0$, as predicted for the global star formation law (equation \ref{eq:globalsf}), the scatter remains nearly the same, a factor of 3.0. Thus we see that the global star formation law is generally a good fit to the extragalactic data. However, if we attempt to include the Galactic observations, the fit severely degrades. The best-fit slope becomes $q=0.48$, with a factor of 21 scatter. Thus the global star formation law cannot fit the Galactic data.

In contrast, if we fit equation (\ref{eq:sflawproj}) to the extragalactic and Galactic data shown in Figure \ref{fig:sigmal}, treating $\epsff$ as a free parameter, we obtain a best-fit value $\epsff=0.010$, with only a factor of 2.8 scatter. Thus the scatter is comparable to that obtained by fitting to the extragalactic data alone in equation (\ref{eq:tdynfit}), but we have now included both the Galactic and the extragalactic data. We therefore conclude that the volumetric star formation law provides a superior match to the data. In fact, the true scatter is probably even smaller than our estimate, because some of the \citet{kennicutt98a} normal disk galaxies that lie below the best-fit line in Figure \ref{fig:sigmal} likely do so because their H$_2$ fractions are small \citep{krumholz09b}, and we have not accounted for this effect as we have in the THINGS data.

\section{Discussion and Conclusion}
\label{sec:discussion}

\subsection{Multiple Star Formation Laws? A Global Law? Thresholds?}
\label{sec:multiple}

\begin{figure}
\epsscale{1.18}
\plotone{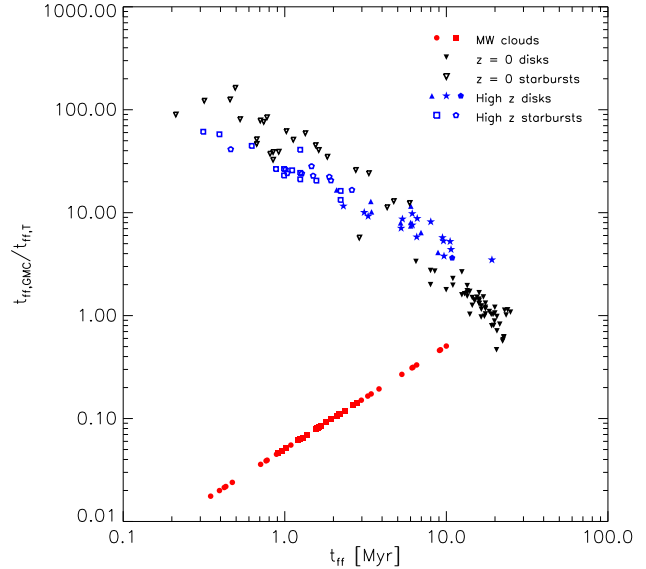}
\caption{
\label{fig:tff}
Ratio of GMC and Toomre free-fall times $\tffgmc/\tfft$ versus minimum free-fall time $\tff=\min(\tffgmc,\tfft)$. All symbols as are in Figure \ref{fig:sigmasigma}. We compute $\tffgmc$ and $\tfft$ for the extragalactic data using equations (\ref{eq:tffexgal1}) and (\ref{eq:tffexgal2}), respectively. For the Milky Way clouds, we take $\tffgmc$ to be equal to the cloud free-fall time, and we compute $\tfft$ using $\Omega = 2\pi / (220\mbox{ Myr})$, the angular velocity at the Solar circle. Because $\tfft$ is the same for all Milky Way clouds, they fall along a line of slope unity. For the Milky Way clouds, note that $\tffgmc$ can be much smaller than for entire disk galaxies because the sample reaches densities of $\sim 10^4$ cm$^{-3}$, a factor of $\sim 100$ denser than the mean GMC density in local disk galaxies. In contrast, the starbursts all have $\tffgmc \gg \tfft$, the $z>0$ disks have $\tffgmc \ga \tfft$, and the $z=0$ disks and Milky Way clouds have $\tffgmc \la \tfft$.
}
\end{figure}


By combining observations of star formation in Galactic and extragalactic systems, we have addressed two important questions about star formation: (1) is the star formation law local, in the sense that the star formation timescale responds primarily to variations in local gas conditions, or global, in the sense that the star formation timescale responds to variations in the galactic orbital period or other galaxy-scale properties? (2) Is the star formation law the same in all molecular clouds, or is there evidence that some clouds obey a different star formation law than others, either because they are in a different galactic environment, or because they are above or below some volume or surface density threshold? We find that a combined Galactic and extragalactic data set favors a local, universal star formation law in which molecular clouds convert their mass into stars at a rate of $\sim 1\%$ of the mass per free-fall time, independent of galactic environment or relationship to any density threshold. 

This is not to say that global galactic properties like the orbital period never have an impact on how stars form in a galaxy. Even though the star formation law is local, galaxy-scale properties like the orbital period can influence star formation if they change the local properties of star-forming molecular clouds. This does not happen in low surface density galaxies like the Milky Way (what we call the GMC regime), as implied by the observations of the Milky Way and the Local Group. However, in galaxies with sufficiently high surface densities (the Toomre regime), which includes almost all mergers and many non-merging high-redshift disks, the weight of the ISM is sufficient to compress molecular clouds to high densities. This alters the local free-fall time within them and thereby raises the star formation rate. Equation (\ref{eq:tffexgal2}) represents our rough attempt to capture this process. Obviously it is a crude approximation, and omits some of the complex physical processes that must take place in a merging or violently gravitationally unstable system, such as compressions produced by galaxy-scale shocks or inflows \citep{barnes04a, saitoh10a, teyssier10a, powell11a}. Nonetheless, we do seem to capture the basic effect, as indicated by the good fit we obtain in Figure \ref{fig:sigmal}.


Similarly, our results do not imply that the SFR in a given molecular cloud is independent of its column density distribution. Indeed, \citet{lada10a} show that the SFR per unit molecular mass in a given cloud is well-correlated with the fraction of the cloud's mass above a K-magnitude extinction of 0.8. The scatter is in SFR per unit mass above $A_K = 0.8$ mag is roughly a factor of $\sim 2$, compared to a factor of $\sim 5$ scatter if one considers all the material above $A_K = 0.1$ mag. This clearly indicates that there is a correlation between the SFR per unit molecular mass and the fraction of a cloud's mass at extinctions above $A_K = 0.8$ mag. However, \citeauthor{lada10a} do not present any evidence that star formation does not occur at column densities below $A_K = 0.8$ mag, and in fact $\sim 2/3$ of the YSOs in the clouds surveyed occur in regions of lower extinction (C. Lada, 2011, private communication). We find here that the Galactic data fall on the same $\dot{\Sigma}_* - \Sigma/\tff$ relation as the extragalactic data, indicating that the SFR per unit mass is also inversely correlated with the free-fall time, regardless of whether one considers the material at $A_K = 0.1$ mag or $A_K = 0.8$ mag.

It is easy to understand why SFR per unit mass correlates with both high extinction and free-fall time. The column density and free-fall time are themselves correlated, in exactly the manner one might have guessed: the clouds with the most mass at high column density are also the ones with the highest volume density, and thus the shortest free-fall time. Thus a correlation between SFR per unit mass and free-fall time implies a correlation between SFR per unit mass and column density distribution, and vice versa. The only question is which correlation is the fundamental one. By themselves just the Milky Way data do not distinguish between these possibilities, and it is possible that both are true to some extent. However, only the free-fall time explanation is able to explain the extragalactic data, and, as we discuss in Section \ref{sec:molcorrelations}, independent lines of evidence from molecular line observations. Thus the most likely explanation for the correlation between SFR and mass at high column density is that column density is correlated with volume density, and not that there is a column density threshold.

\subsection{The Disk-Starburst Bimodality}

Since the star formation law is universal, how can we then explain the apparent bimodality between disks and starbursts seen in Figure \ref{fig:sigmasigma}, or in similar plots of $L_{\rm IR}$ versus $M_{\rm H_2}$ (e.g.~Figure 1 of \citealt{daddi10a}, or Figure 2 of \citealt{genzel10a})? Part of the answer is that the bimodality is artificially enhanced by two effects. One is the use of a CO-H$_2$ conversion factor that jumps discontinuously between disk and starburst galaxies, rather than varying continuously with galaxy parameters. This is probably a significant oversimplification \citep{narayanan11b}. The second is selection bias, with starburst galaxies being selected based on the extremity of their properties, for example their extremely high sub-mm fluxes. Such selection preferentially picks out objects that are as far as possible from the ``normal" star-forming galaxy sequence. In the unbiased sample provided by the COLDGASS survey \citep{saintonge11a, saintonge11b}, the H$_2$ depletion time, defined as \begin{equation}
t_{\rm dep,H_2} = \frac{M_{\rm H_2}}{\dot{M}_*},
\end{equation}
is continuously rather than bimodally distributed, and the data occupy the full range of values between the sequences of disks and starbursts identified by \citet{daddi10a} and \citet{genzel10a}.

Even if the combined effects of the CO-H$_2$ conversion factor and selection bias explain the bimodality, however, there remains the question of why the Local Group galaxies show H$_2$ depletion times with a nearly constant value $t_{\rm dep,H_2} \approx 2$ Gyr (\citealt{bigiel08a, leroy08a}; also see \citealt{young89a}, who report constant depletion times over a larger sample of unresolved galaxies) while in samples that include either local molecular clouds or a broader range of galaxies, whether at $z=0$ or at high redshift, $t_{\rm dep,H_2}$ is not constant. The answer can be found in how the density and free-fall time in star-forming gas clouds depends, or does not depend, on global galactic quantities. Figure \ref{fig:tff} shows the ratio of $\tffgmc$ to $\tfft$ for both the unresolved galaxies and the Galactic clouds in our sample. The plot immediately shows an important dichotomy. The starburst galaxies have $\tfft \ll \tffgmc$, the high-$z$ disks have $\tfft \la \tffgmc$, but the local galaxies and Milky Way clouds have $\tfft \ga \tffgmc$. This is exactly as we expect: in galaxies like the Milky Way, GMCs are overdense, bound objects that decouple from the rest of the ISM. As a result, the free-fall times in these objects are set by their internal properties and processes, and not by the large-scale behavior of the ISM.

For the clouds in the \citet{heiderman10a} and \citet{lada10a} $A_K = 0.8$ mag samples, this effect is particularly pronounced, because the depletion time is
\begin{equation}
t_{\rm dep,H_2} = \frac{\tff}{\epsff} = 0.43 \epsilon_{\rm ff,-2}^{-1} n_2^{-1/2}\mbox{ Gyr},
\end{equation}
where $\epsilon_{\rm ff,-2} = \epsff/100$, $n_2 = n/100$ H nuclei cm$^{-3}$, and we have used a mean mass per H nucleus of $\mu_{\rm H} = 2.3\times 10^{-24}$ g. The mean densities of the clouds in these samples are $10^3 - 10^4$ cm$^{-3}$, compared to the ISM average of 1 cm$^{-3}$, and to an average of $\sim 30$ cm$^{-3}$ for all the molecular gas in the Galaxy \citep{mckee99a}. This means that their depletion times are far smaller than the average even over Local Group galaxies, simply because their free-fall times are also much smaller than the mean of the molecular gas in these galaxies. Even for entire galaxies that fall into the regime where $\tfft > \tffgmc$, however, the depletion time is
\begin{eqnarray}
t_{\rm dep,H_2} & = & \frac{\tffgmc}{\epsff} = \frac{\pi^{1/4}}{\sqrt{8}} \frac{\sigma}{\epsff G(\Sigma_{\rm GMC}^3 \Sigma_{\rm gal})^{1/4}}
\nonumber \\
& = & 1.9\epsilon_{\rm ff,-2}^{-1} \sigma_1 \Sigma_{\rm GMC,2}^{-3/4} \Sigma_{\rm gal,1}^{-1/4}\mbox{ Gyr},
\end{eqnarray}
where $\sigma_1 = \sigma/10$ km s$^{-1}$, $\Sigma_{\rm GMC,2} = \Sigma_{\rm GMC}/100$ $\msun$ pc$^{-2}$, and $\Sigma_{\rm gal,1} = \Sigma_{\rm gal}/10$ $\msun$ pc$^{-2}$. Observations indicate that $\epsff$, $\sigma$, and $\Sigma_{\rm GMC}$ are essentially invariant across the range of galaxies sampled by THINGS, which includes only quiescent objects (not mergers) at redshift 0. The only quantity that does vary, $\Sigma_{\rm gal}$, enters with a $-1/4$ power dependence. This is why $t_{\rm dep,H_2}$ is observed to be essentially invariant at a value of $\sim 2$ Gyr across the THINGS sample. It is interesting to note that the density corresponding to this depletion time is $n\sim 5$ cm$^{-3}$, lower that the typical observed GMC density in the Solar neighborhood. However, recall that most of the molecular mass in a galaxy is in the most massive GMCs \citep{rosolowsky05b}, and that, at fixed surface density, the volume density varies with GMC mass as $M_{\rm GMC}^{-1/2}$. Thus we naturally obtain lower volume densities for the bulk of the mass, although our density estimate is probably somewhat too low, since the values of $M_{\rm GMC}$ we obtain tend to be characteristic of the largest GMCs in a galaxy, rather than the median.

In contrast, in starbursts and high-redshift galaxies, star-forming regions are not able to decouple from the ambient ISM, and wind up being only mildly overdense. As a result their free-fall times are set by the large-scale properties of the ISM, and $\tffgmc > \tfft$. These galaxies have depletion times
\begin{eqnarray}
t_{\rm dep,H_2} & = & \frac{\tfft}{\epsff} = \sqrt{\frac{3\pi^4 Q^2}{8(\beta+1)\phi_P}} \frac{\torb}{\epsff}
\nonumber \\
& = & 350 \epsilon_{\rm ff,-2}^{-1} Q \torb,
\end{eqnarray}
where for the numerical evaluation we have used $\beta=0$ and $\phi_P = 3$. Thus the depletion time scales linearly with the orbital period in the Toomre regime where $\tffgmc > \tfft$. We can therefore understand why the depletion time is not constant in broader galaxy samples than those limited to the Local Group. In these broader samples, some or all of the galaxies are in the Toomre regime, where $\torb$, which matters, varies strongly with redshift and depends on whether a galaxy is quiescently forming stars or a starburst. In contrast, the bulk of disk galaxies at $z=0$ are in the GMC regime, where $t_{\rm orb}$ does not matter.

\subsection{Relationship to Molecular Line - Star Formation Correlations}
\label{sec:molcorrelations}

The data set we have gathered here represents one approach to the problem of determining the star formation law: combining spatially resolved and unresolved observations of the correlation between star formation and the bulk of molecular gas. An orthogonal approach is to use solely unresolved observations, but to measure the correlation between the star formation rate and the luminosity in a wide variety of molecular lines.
Studies based on this approach include \citet{gao04a}, \citet{wu05a, wu10a}, \citet{narayanan08a}, \citet{bussmann08a}, \citet{bayet09a} \citet{juneau09a}, and \citet{schenck11a}. Since different lines provide information about gas at different densities, the use of multiple molecular lines provides density resolution akin to the spatial resolution we obtain here by including the Galactic and Local Group data together with the unresolved observations. 

We first note that a general result of these surveys is that, in the most rapidly star-forming galaxies, a significant fraction of the ISM mass can reside at the densities $\sim 10^4$ cm$^{-3}$ or more traced by lines like HCN($1\rightarrow 0$). This is consistent with the results shown in Figure \ref{fig:sigmasigma}, which indicates that, if we adopt the local star formation law, the most strongly star-forming galaxies must have mean volume densities in this range. To first approximation these high volume densities can be understood as the result of the requirements of vertical pressure balance and marginal gravitational stability in a high surface-density disk, effects captured in Equation (\ref{eq:tffexgal2}); as noted in Section \ref{sec:multiple}, additional processes that we have not modeled may also play a role.

Turning to a more quantitative analysis of the molecular line observations, one early result was that the correlation between the far infrared and HCN($1\rightarrow 0$) line luminosities of galaxies is close to linear \citep{gao04a, wu05a, wu10a}. This linearity was one of the original motivations for the threshold model. However, subsequent work has shown that the relation deviates from linearity at very high infrared luminosity \citep{gao07a}, and that lines with critical densities higher than HCN($1\rightarrow 0$) generally show sub-linear FIR-line correlations \citep[e.g.][]{narayanan08a, bussmann08a, bayet09a, juneau09a}, suggesting that the situation is somewhat more complex.

To date the only published theoretical models for the molecular line-star formation rate correlation are those of \citet{krumholz07g} and \citet{narayanan08b}, who adopt a model for star formation equivalent to the local, volumetric star formation law (equation \ref{eq:sflaw}). The primary result of this work is a prediction that the powerlaw index $p$ in the star formation-molecular line correlation $\dot{M}_* \propto L_{\rm line}^p$ should depend on the ratio of the mean density $n$ in a galaxy to the critical density $n_{\rm crit}$ of the molecule being observed. For $n\gg n_{\rm crit}$, as one expects for low critical density transitions such as CO($1\rightarrow 0$) or in very high density galaxies like ULIRGs, the observation probes the entire mass of the ISM, and one should have $p\approx 1.5$. For $n\ltsim n_{\rm crit}$, as is expected for high critical density transitions like HCN($1\rightarrow 0$) in normal galaxies, the index $p$ decreases, approaching unity. The index $p$ can even fall below unity for $n$ sufficiently small compared to $n_{\rm crit}$, for example in the case of the HCN($3\rightarrow 2$) transition \citep{bussmann08a}. More precise and quantitative predictions for various molecules are given in \citeauthor{krumholz07g} and \citeauthor{narayanan08b}

Thus far observations show very good agreement with these models \citep[e.g.][]{narayanan08a, bussmann08a, bayet09a, juneau09a, schenck11a}. In particular, the observations confirm the prediction that transitions with sufficiently high critical densities give rise to values of $p < 1$ (\citealt{bussmann08a} and \citealt{juneau09a} for HCN($3\rightarrow 2$); \citealt{bayet09a} for CO($J+1\rightarrow J$) with $J>5$). In contrast, \citet{wu10a} report that they do not find good agreement with the \citeauthor{krumholz07g} and \citeauthor{narayanan08b} models in a survey of Galactic sources. However, as noted in Section \ref{sec:galdata}, these observations are certainly compromised by the fact that the regions observed by \citeauthor{wu10a}\ are too young for the infrared luminosity to serve as a reasonable proxy for the star formation rate, as \citeauthor{wu10a}\ assume; indeed, \citet{heiderman10a} note that the \citeauthor{wu10a}'s assumption likely introduces an order of magnitude-level systematic error.

No comparable predictions exist for the global star formation law, equation (\ref{eq:globalsf}), so it is not clear whether these models will be able to explain the observations. For the threshold models, on the other hand, the molecular line observations present another clear problem. If the star formation rate is simply the mass of dense gas divided by a constant star formation timescale $t_{\rm dense}$, then, as \citet{bussmann08a} point out, the star formation rate should simply correlate linearly with the mass of gas above the density threshold. Thus we would expect $p > 1$ for any transition where $n_{\rm crit}$ is below the star formation density threshold, and $p = 1$ for transitions where $n_{\rm crit}$ is well above the density threshold. Values of $p<1$ should be impossible. Thus we see that the assumption of a constant $t_{\rm dense}$ in the threshold model is also inconsistent with the molecular line observations.

Thus our finding that a local, volumetric star formation law provides the best fit to the combination of Galactic, Local Group, and unresolved extragalactic observations of the molecular gas-star formation correlation, while the threshold model does not, is consistent with the results of comparing theoretical models to the observed line luminosity-star formation correlation.

\subsection{Implications}

Our conclusion that the underlying physics of star formation obeys a simple, local, volumetric law has several important implications. First, it validates the use of the local star formation law, equation (\ref{eq:sflaw}) with $\epsff \approx 0.01$, as one of the standard recipes in numerical simulations of star formation on galactic or cosmological scales \citep[e.g.][]{springel03a, robertson08b, gnedin09a, bournaud10a, teyssier10a, ceverino10a, agertz11a, kuhlen11a}, and suggests that there is no need to modify these laws to contain additional factors that depend on the bulk properties of galaxies. Nor is it necessary that the simulations have resolution sufficient to exceed the proposed high density threshold of $\sim 10^4-10^5$ cm$^{-3}$.\footnote{As noted above, it is important to distinguish thresholds of $\sim 10^4-10^5$ cm$^{-3}$ that apply in purely molecular gas from thresholds of $\sim 1-10$ cm$^{-3}$ that are used to separate atomic from molecular gas in codes that do not include an explicit treatment of molecule formation. The latter is a means of approximating a real change in the physical state of the ISM that does affect how stars form \citep{krumholz11b}, while there is no change in the physical state of the ISM or of the star formation process associated with the former.
} 
In order to trust the star formation rate in a simulated galaxy, one must still resolve the mean density in its star-forming clouds by a safe margin.
 Note that resolving a given density means not just that it is possible for the gas to reach that density in a simulation, but that the resolution is high enough that the behavior of the gas is not be compromised by artificial pressure floors, artificial fragmentation, or other numerical artifacts. This condition is necessary in order to obtain an accurate estimate of $\tff$ in equation (\ref{eq:sflaw}). Simulations that fail to do so can underestimate the star formation rate \citep{teyssier10a}. However, once this goal is achieved, it is not necessary to go further and resolve the extreme tail of the density PDF that extends above $\sim 10^4-10^5$ cm$^{-3}$. Indeed, our conclusion is consistent with the numerical simulations of \citet{teyssier10a}, who find from high-resolution simulations of mergers that the difference between the disk and starburst star formation laws proposed by \citet{daddi10a} can be fully accounted for simply by an increase in the mean ISM density in starbursts, which produces a corresponding decrease in the free-fall time $\tff$. Both the functional form of the star formation law (equation \ref{eq:sflaw}) and the value of $\epsff$, the fraction of the mass transformed into stars per free-fall time, remain unchanged, and a star formation threshold of $\sim 10$ cm$^{-3}$ is sufficient. 
 
 What density resolution these considerations imply in practice will vary depending on what details it is important that a given simulation gets right. If the goal is to compare mergers and disks, as in \citet{teyssier10a}, one must clearly resolve the mean ISM density of $\sim 10^4$ cm$^{-3}$ in the mergers. In large-volume cosmological simulations where one is mainly concerned with mean properties of large numbers of galaxies rather than the star formation law within individual galaxies, a lower density resolution is probably acceptable, 
although we note that even in Milky Way-like galaxies the mean molecular cloud density is $\sim 10^2$ cm$^{-3}$, and thus one will only obtain an accurate estimate of $\tff$ if the resolution is high enough for the physics at this density to be trustworthy.
Few cosmological, or even isolated galaxy, simulations achieve this goal. Finally, we note that our discussion does not address the issue of what density resolution is required for a correct treatment of star formation {\it feedback} \citep[e.g.][]{governato10a, brook11a}. This need not be the same as the values quoted above, since the physics that governs, e.g., the interaction of supernova blast waves with a clumpy ISM is quite different than that which regulates star formation in cold molecular clouds.

Second, our conclusion implies that the star formation rate in galaxies cannot solely be determined by feedback produced by massive stars (supernovae, stellar winds) as hypothesized by several authors \citep{dekel86a, murray10a, dobbs11a, hopkins11a}. This feedback is undoubtedly important, and must be included in simulations if one wishes to obtain realistic values for quantities like the galactic scale height or the mass in different ISM phases. However, if massive stars were the only mechanism at work there would be no reason for small molecular clouds lacking in massive stars, such as the majority of those found in the \citet{lada10a} and \citet{heiderman10a} samples, to lie on the extragalactic star formation law. Instead, since they lack massive star feedback, one would have expected these systems to show significantly higher values of $\epsff$ than extragalactic systems. Instead, the value of $\epsff$ appears to be independent of the presence or absence of massive stars. One possible explanation for the invariance of $\epsff$ with the presence of absence of massive stars is that turbulence regulates the SFR \citep{krumholz05c}, since the properties of the turbulence will be largely independent of the exact mechanism by which it is driven. In regions lacking massive stars the turbulence can be driven by mechanisms such as protostellar outflows \citep{li06b, nakamura07a, matzner07a}, while in regions containing massive stars it is driven by the radiation pressure \citep{krumholz09d, murray10a} or supernovae. Regardless of what mechanism is responsible for setting it, however, the observations clearly show that the value of $\epsff$ is roughly constant in star-forming systems from nearby low mass clouds to entire starburst galaxies, as demonstrated in Figure \ref{fig:sigmal}.

\acknowledgements 
We  thank E.~Daddi, R.~Genzel, D.~Narayanan, S.~Oey, E.~Ostriker, N.~Scoville, and R.~Teyssier for helpful discussions and comments on the manuscript, F.~Bigiel, E.~Daddi, C.~Lada, and A.~Leroy for providing copies of their data and assistance in working with it, and F.~Bournaud for a helpful and timely referee report. We acknowledge support from: an Alfred P.~Sloan Fellowship (MRK); the National Science Foundation through grants AST-0807739 (MRK), AST-0908553 (CFM), AST-1010033 (AD), and CAREER-0955300 (MRK); NASA through Astrophysics Theory and Fundamental Physics grant NNX09AK31G (CFM and MRK) and a {\it Chandra Space Telescope} grant (MRK); the ISF through grant 6/08 (AD); the GIF through grant G-1052-104.7/2009 (AD); and a DIP grant (AD).

\bibliographystyle{apj}
\bibliography{refs}

\begin{thebibliography}{101}
\expandafter\ifx\csname natexlab\endcsname\relax\def\natexlab#1{#1}\fi

\bibitem[{{Agertz} {et~al.}(2011){Agertz}, {Teyssier}, \& {Moore}}]{agertz11a}
{Agertz}, O., {Teyssier}, R., \& {Moore}, B. 2011, \mnras, 410, 1391

\bibitem[{{Barnes}(2004)}]{barnes04a}
{Barnes}, J.~E. 2004, \mnras, 350, 798

\bibitem[{{Bayet} {et~al.}(2009){Bayet}, {Gerin}, {Phillips}, \&
  {Contursi}}]{bayet09a}
{Bayet}, E., {Gerin}, M., {Phillips}, T.~G., \& {Contursi}, A. 2009, \mnras,
  399, 264

\bibitem[{{Bigiel} {et~al.}(2008){Bigiel}, {Leroy}, {Walter}, {Brinks}, {de
  Blok}, {Madore}, \& {Thornley}}]{bigiel08a}
{Bigiel}, F., {Leroy}, A., {Walter}, F., {Brinks}, E., {de Blok}, W.~J.~G.,
  {Madore}, B., \& {Thornley}, M.~D. 2008, \aj, 136, 2846

\bibitem[{{Blanc} {et~al.}(2009){Blanc}, {Heiderman}, {Gebhardt}, {Evans}, \&
  {Adams}}]{blanc09a}
{Blanc}, G.~A., {Heiderman}, A., {Gebhardt}, K., {Evans}, N.~J., \& {Adams}, J.
  2009, \apj, 704, 842

\bibitem[{{Bolatto} {et~al.}(2008){Bolatto}, {Leroy}, {Rosolowsky}, {Walter},
  \& {Blitz}}]{bolatto08a}
{Bolatto}, A.~D., {Leroy}, A.~K., {Rosolowsky}, E., {Walter}, F., \& {Blitz},
  L. 2008, \apj, 686, 948

\bibitem[{{Bouch{\'e}} {et~al.}(2007){Bouch{\'e}}, {Cresci}, {Davies},
  {Eisenhauer}, {F{\"o}rster Schreiber}, {Genzel}, {Gillessen}, {Lehnert},
  {Lutz}, {Nesvadba}, {Shapiro}, {Sternberg}, {Tacconi}, {Verma}, {Cimatti},
  {Daddi}, {Renzini}, {Erb}, {Shapley}, \& {Steidel}}]{bouche07a}
{Bouch{\'e}}, N., {et~al.} 2007, \apj, 671, 303

\bibitem[{{Boulares} \& {Cox}(1990)}]{boulares90a}
{Boulares}, A., \& {Cox}, D.~P. 1990, \apj, 365, 544

\bibitem[{{Bournaud} {et~al.}(2010){Bournaud}, {Elmegreen}, {Teyssier},
  {Block}, \& {Puerari}}]{bournaud10a}
{Bournaud}, F., {Elmegreen}, B.~G., {Teyssier}, R., {Block}, D.~L., \&
  {Puerari}, I. 2010, \mnras, 409, 1088

\bibitem[{{Brook} {et~al.}(2011){Brook}, {Governato}, {Ro{\v s}kar}, {Stinson},
  {Brooks}, {Wadsley}, {Quinn}, {Gibson}, {Snaith}, {Pilkington}, {House}, \&
  {Pontzen}}]{brook11a}
{Brook}, C.~B., {et~al.} 2011, \mnras, 595

\bibitem[{{Bussmann} {et~al.}(2008){Bussmann}, {Narayanan}, {Shirley},
  {Juneau}, {Wu}, {Solomon}, {Vanden Bout}, {Moustakas}, \&
  {Walker}}]{bussmann08a}
{Bussmann}, R.~S., {et~al.} 2008, \apjl, 681, L73

\bibitem[{{Ceverino} {et~al.}(2010){Ceverino}, {Dekel}, \&
  {Bournaud}}]{ceverino10a}
{Ceverino}, D., {Dekel}, A., \& {Bournaud}, F. 2010, \mnras, 404, 2151

\bibitem[{{Ceverino} {et~al.}(2011){Ceverino}, {Dekel}, {Mandelker},
  {Bournaud}, {Burkert}, {Genzel}, \& {Primack}}]{ceverino11a}
{Ceverino}, D., {Dekel}, A., {Mandelker}, N., {Bournaud}, F., {Burkert}, A.,
  {Genzel}, R., \& {Primack}, J. 2011, ArXiv e-prints

\bibitem[{{Chung} {et~al.}(2009){Chung}, {van Gorkom}, {Kenney}, {Crowl}, \&
  {Vollmer}}]{chung09a}
{Chung}, A., {van Gorkom}, J.~H., {Kenney}, J.~D.~P., {Crowl}, H., \&
  {Vollmer}, B. 2009, \aj, 138, 1741

\bibitem[{{Cresci} {et~al.}(2009){Cresci}, {Hicks}, {Genzel}, {Schreiber},
  {Davies}, {Bouch{\'e}}, {Buschkamp}, {Genel}, {Shapiro}, {Tacconi},
  {Sommer-Larsen}, {Burkert}, {Eisenhauer}, {Gerhard}, {Lutz}, {Naab},
  {Sternberg}, {Cimatti}, {Daddi}, {Erb}, {Kurk}, {Lilly}, {Renzini},
  {Shapley}, {Steidel}, \& {Caputi}}]{cresci09a}
{Cresci}, G., {et~al.} 2009, \apj, 697, 115

\bibitem[{{Daddi} {et~al.}(2010{\natexlab{a}}){Daddi}, {Bournaud}, {Walter},
  {Dannerbauer}, {Carilli}, {Dickinson}, {Elbaz}, {Morrison}, {Riechers},
  {Onodera}, {Salmi}, {Krips}, \& {Stern}}]{daddi10b}
{Daddi}, E., {et~al.} 2010{\natexlab{a}}, \apj, 713, 686

\bibitem[{{Daddi} {et~al.}(2008){Daddi}, {Dannerbauer}, {Elbaz}, {Dickinson},
  {Morrison}, {Stern}, \& {Ravindranath}}]{daddi08a}
{Daddi}, E., {Dannerbauer}, H., {Elbaz}, D., {Dickinson}, M., {Morrison}, G.,
  {Stern}, D., \& {Ravindranath}, S. 2008, \apjl, 673, L21

\bibitem[{{Daddi} {et~al.}(2010{\natexlab{b}}){Daddi}, {Elbaz}, {Walter},
  {Bournaud}, {Salmi}, {Carilli}, {Dannerbauer}, {Dickinson}, {Monaco}, \&
  {Riechers}}]{daddi10a}
{Daddi}, E., {et~al.} 2010{\natexlab{b}}, \apjl, 714, L118

\bibitem[{{Dekel} {et~al.}(2009{\natexlab{a}}){Dekel}, {Birnboim}, {Engel},
  {Freundlich}, {Goerdt}, {Mumcuoglu}, {Neistein}, {Pichon}, {Teyssier}, \&
  {Zinger}}]{dekel09b}
{Dekel}, A., {et~al.} 2009{\natexlab{a}}, \nat, 457, 451

\bibitem[{{Dekel} {et~al.}(2009{\natexlab{b}}){Dekel}, {Sari}, \&
  {Ceverino}}]{dekel09a}
{Dekel}, A., {Sari}, R., \& {Ceverino}, D. 2009{\natexlab{b}}, \apj, 703, 785

\bibitem[{{Dekel} \& {Silk}(1986)}]{dekel86a}
{Dekel}, A., \& {Silk}, J. 1986, \apj, 303, 39

\bibitem[{{Dib} {et~al.}(2006){Dib}, {Bell}, \& {Burkert}}]{dib06a}
{Dib}, S., {Bell}, E., \& {Burkert}, A. 2006, \apj, 638, 797

\bibitem[{{Dobbs} {et~al.}(2011){Dobbs}, {Burkert}, \& {Pringle}}]{dobbs11a}
{Dobbs}, C.~L., {Burkert}, A., \& {Pringle}, J.~E. 2011, \mnras, in press,
  arXiv:1107.0154

\bibitem[{{Downes} \& {Solomon}(1998)}]{downes98a}
{Downes}, D., \& {Solomon}, P.~M. 1998, \apj, 507, 615

\bibitem[{{Elmegreen}(2002)}]{elmegreen02a}
{Elmegreen}, B.~G. 2002, \apj, 577, 206

\bibitem[{{Evans} {et~al.}(2009){Evans}, {Dunham}, {J{\o}rgensen}, {Enoch},
  {Mer{\'{\i}}n}, {van Dishoeck}, {Alcal{\'a}}, {Myers}, {Stapelfeldt},
  {Huard}, {Allen}, {Harvey}, {van Kempen}, {Blake}, {Koerner}, {Mundy},
  {Padgett}, \& {Sargent}}]{evans09a}
{Evans}, N.~J., {et~al.} 2009, \apjs, 181, 321

\bibitem[{{Fich} {et~al.}(1989){Fich}, {Blitz}, \& {Stark}}]{fich89a}
{Fich}, M., {Blitz}, L., \& {Stark}, A.~A. 1989, \apj, 342, 272

\bibitem[{{Fukui} \& {Kawamura}(2010)}]{fukui10a}
{Fukui}, Y., \& {Kawamura}, A. 2010, \araa, 48, 547

\bibitem[{{Fukui} {et~al.}(2009){Fukui}, {Kawamura}, {Wong}, {Murai},
  {Iritani}, {Mizuno}, {Mizuno}, {Onishi}, {Hughes}, {Ott}, {Muller},
  {Staveley-Smith}, \& {Kim}}]{fukui09a}
{Fukui}, Y., {et~al.} 2009, \apj, 705, 144

\bibitem[{{Gao} {et~al.}(2007){Gao}, {Carilli}, {Solomon}, \& {Vanden
  Bout}}]{gao07a}
{Gao}, Y., {Carilli}, C.~L., {Solomon}, P.~M., \& {Vanden Bout}, P.~A. 2007,
  \apjl, 660, L93

\bibitem[{{Gao} \& {Solomon}(2004)}]{gao04a}
{Gao}, Y., \& {Solomon}, P.~M. 2004, \apj, 606, 271

\bibitem[{{Genzel} {et~al.}(2010){Genzel}, {Tacconi}, {Gracia-Carpio},
  {Sternberg}, {Cooper}, {Shapiro}, {Bolatto}, {Bouch{\'e}}, {Bournaud},
  {Burkert}, {Combes}, {Comerford}, {Cox}, {Davis}, {Schreiber},
  {Garcia-Burillo}, {Lutz}, {Naab}, {Neri}, {Omont}, {Shapley}, \&
  {Weiner}}]{genzel10a}
{Genzel}, R., {et~al.} 2010, \mnras, 407, 2091

\bibitem[{{Ghez} {et~al.}(2008){Ghez}, {Salim}, {Weinberg}, {Lu}, {Do}, {Dunn},
  {Matthews}, {Morris}, {Yelda}, {Becklin}, {Kremenek}, {Milosavljevic}, \&
  {Naiman}}]{ghez08a}
{Ghez}, A.~M., {et~al.} 2008, \apj, 689, 1044

\bibitem[{{Gillessen} {et~al.}(2009){Gillessen}, {Eisenhauer}, {Fritz},
  {Bartko}, {Dodds-Eden}, {Pfuhl}, {Ott}, \& {Genzel}}]{gillessen09a}
{Gillessen}, S., {Eisenhauer}, F., {Fritz}, T.~K., {Bartko}, H., {Dodds-Eden},
  K., {Pfuhl}, O., {Ott}, T., \& {Genzel}, R. 2009, \apjl, 707, L114

\bibitem[{{Gnedin} \& {Kravtsov}(2010)}]{gnedin10a}
{Gnedin}, N.~Y., \& {Kravtsov}, A.~V. 2010, \apj, 714, 287

\bibitem[{{Gnedin} {et~al.}(2009){Gnedin}, {Tassis}, \& {Kravtsov}}]{gnedin09a}
{Gnedin}, N.~Y., {Tassis}, K., \& {Kravtsov}, A.~V. 2009, \apj, 697, 55

\bibitem[{{Goldbaum} {et~al.}(2011){Goldbaum}, {Krumholz}, {Matzner}, \&
  {McKee}}]{goldbaum11a}
{Goldbaum}, N.~J., {Krumholz}, M.~R., {Matzner}, C.~D., \& {McKee}, C.~F. 2011,
  \apj, 738, 101

\bibitem[{{Governato} {et~al.}(2010){Governato}, {Brook}, {Mayer}, {Brooks},
  {Rhee}, {Wadsley}, {Jonsson}, {Willman}, {Stinson}, {Quinn}, \&
  {Madau}}]{governato10a}
{Governato}, F., {et~al.} 2010, \nat, 463, 203

\bibitem[{{Heiderman} {et~al.}(2010){Heiderman}, {Evans}, {Allen}, {Huard}, \&
  {Heyer}}]{heiderman10a}
{Heiderman}, A., {Evans}, II, N.~J., {Allen}, L.~E., {Huard}, T., \& {Heyer},
  M. 2010, \apj, 723, 1019

\bibitem[{{Hopkins} {et~al.}(2011){Hopkins}, {Quataert}, \&
  {Murray}}]{hopkins11a}
{Hopkins}, P.~F., {Quataert}, E., \& {Murray}, N. 2011, \mnras, 1513, in press,
  arXiv:1101.4940

\bibitem[{{Juneau} {et~al.}(2009){Juneau}, {Narayanan}, {Moustakas}, {Shirley},
  {Bussmann}, {Kennicutt}, \& {Vanden Bout}}]{juneau09a}
{Juneau}, S., {Narayanan}, D.~T., {Moustakas}, J., {Shirley}, Y.~L.,
  {Bussmann}, R.~S., {Kennicutt}, Jr., R.~C., \& {Vanden Bout}, P.~A. 2009,
  \apj, 707, 1217

\bibitem[{{Kennicutt}(1989)}]{kennicutt89b}
{Kennicutt}, R.~C. 1989, \apj, 344, 685

\bibitem[{{Kennicutt}(1998)}]{kennicutt98a}
{Kennicutt}, Jr., R.~C. 1998, \apj, 498, 541

\bibitem[{{Kennicutt} {et~al.}(2007){Kennicutt}, {Calzetti}, {Walter}, {Helou},
  {Hollenbach}, {Armus}, {Bendo}, {Dale}, {Draine}, {Engelbracht}, {Gordon},
  {Prescott}, {Regan}, {Thornley}, {Bot}, {Brinks}, {de Blok}, {de Mello},
  {Meyer}, {Moustakas}, {Murphy}, {Sheth}, \& {Smith}}]{kennicutt07a}
{Kennicutt}, Jr., R.~C., {et~al.} 2007, \apj, 671, 333

\bibitem[{{Kim} {et~al.}(2011){Kim}, {Kim}, \& {Ostriker}}]{kim11a}
{Kim}, C.-G., {Kim}, W.-T., \& {Ostriker}, E.~C. 2011, \apj, in press,
  arXiv:1109.0028

\bibitem[{{Kim} \& {Ostriker}(2002)}]{kim02a}
{Kim}, W., \& {Ostriker}, E.~C. 2002, \apj, 570, 132

\bibitem[{{Krumholz} \& {Dekel}(2011)}]{krumholz11d}
{Krumholz}, M.~R., \& {Dekel}, A. 2011, \apj, submitted, arXiv:1106.0301

\bibitem[{{Krumholz} {et~al.}(2011){Krumholz}, {Leroy}, \&
  {McKee}}]{krumholz11b}
{Krumholz}, M.~R., {Leroy}, A.~K., \& {McKee}, C.~F. 2011, \apj, 731, 25

\bibitem[{{Krumholz} \& {Matzner}(2009)}]{krumholz09d}
{Krumholz}, M.~R., \& {Matzner}, C.~D. 2009, \apj, 703, 1352

\bibitem[{{Krumholz} {et~al.}(2006){Krumholz}, {Matzner}, \&
  {McKee}}]{krumholz06d}
{Krumholz}, M.~R., {Matzner}, C.~D., \& {McKee}, C.~F. 2006, \apj, 653, 361

\bibitem[{{Krumholz} \& {McKee}(2005)}]{krumholz05c}
{Krumholz}, M.~R., \& {McKee}, C.~F. 2005, \apj, 630, 250

\bibitem[{{Krumholz} {et~al.}(2009){Krumholz}, {McKee}, \&
  {Tumlinson}}]{krumholz09b}
{Krumholz}, M.~R., {McKee}, C.~F., \& {Tumlinson}, J. 2009, \apj, 699, 850

\bibitem[{{Krumholz} \& {Tan}(2007)}]{krumholz07e}
{Krumholz}, M.~R., \& {Tan}, J.~C. 2007, \apj, 654, 304

\bibitem[{{Krumholz} \& {Thompson}(2007)}]{krumholz07g}
{Krumholz}, M.~R., \& {Thompson}, T.~A. 2007, \apj, 669, 289

\bibitem[{{Kuhlen} {et~al.}(2011){Kuhlen}, {Krumholz}, {Madau}, {Smith}, \&
  {Wise}}]{kuhlen11a}
{Kuhlen}, M., {Krumholz}, M., {Madau}, P., {Smith}, B., \& {Wise}, J. 2011,
  \apj, submitted, arXiv:1105.2376

\bibitem[{{Lada} {et~al.}(2010){Lada}, {Lombardi}, \& {Alves}}]{lada10a}
{Lada}, C.~J., {Lombardi}, M., \& {Alves}, J.~F. 2010, \apj, 724, 687

\bibitem[{{Leroy} {et~al.}(2009){Leroy}, {Walter}, {Bigiel}, {Usero}, {Weiss},
  {Brinks}, {de Blok}, {Kennicutt}, {Schuster}, {Kramer}, {Wiesemeyer}, \&
  {Roussel}}]{leroy09a}
{Leroy}, A.~K., {et~al.} 2009, \aj, 137, 4670

\bibitem[{{Leroy} {et~al.}(2008){Leroy}, {Walter}, {Brinks}, {Bigiel}, {de
  Blok}, {Madore}, \& {Thornley}}]{leroy08a}
{Leroy}, A.~K., {Walter}, F., {Brinks}, E., {Bigiel}, F., {de Blok}, W.~J.~G.,
  {Madore}, B., \& {Thornley}, M.~D. 2008, \aj, 136, 2782

\bibitem[{{Li} {et~al.}(2005){Li}, {Mac Low}, \& {Klessen}}]{li05a}
{Li}, Y., {Mac Low}, M., \& {Klessen}, R.~S. 2005, \apjl, 620, L19

\bibitem[{{Li} \& {Nakamura}(2006)}]{li06b}
{Li}, Z.-Y., \& {Nakamura}, F. 2006, \apjl, 640, L187

\bibitem[{{Lombardi} {et~al.}(2010){Lombardi}, {Alves}, \&
  {Lada}}]{lombardi10a}
{Lombardi}, M., {Alves}, J., \& {Lada}, C.~J. 2010, \aap, 519, L7+

\bibitem[{{Matzner}(2007)}]{matzner07a}
{Matzner}, C.~D. 2007, \apj, 659, 1394

\bibitem[{{McKee}(1989)}]{mckee89a}
{McKee}, C.~F. 1989, \apj, 345, 782

\bibitem[{{McKee}(1999)}]{mckee99a}
{McKee}, C.~F. 1999, in NATO ASIC Proc. 540: The Origin of Stars and Planetary
  Systems, ed. {C.~J.~Lada \& N.~D.~Kylafis}, 29

\bibitem[{{McKee} \& {Ostriker}(2007)}]{mckee07a}
{McKee}, C.~F., \& {Ostriker}, E.~C. 2007, \araa, 45, 565

\bibitem[{{Mooney} \& {Solomon}(1988)}]{mooney88a}
{Mooney}, T.~J., \& {Solomon}, P.~M. 1988, \apjl, 334, L51

\bibitem[{{Murray} {et~al.}(2010){Murray}, {Quataert}, \&
  {Thompson}}]{murray10a}
{Murray}, N., {Quataert}, E., \& {Thompson}, T.~A. 2010, \apj, 709, 191

\bibitem[{{Nakamura} \& {Li}(2007)}]{nakamura07a}
{Nakamura}, F., \& {Li}, Z.-Y. 2007, \apj, 662, 395

\bibitem[{{Narayanan} {et~al.}(2008{\natexlab{a}}){Narayanan}, {Cox}, \&
  {Hernquist}}]{narayanan08a}
{Narayanan}, D., {Cox}, T.~J., \& {Hernquist}, L. 2008{\natexlab{a}}, \apjl,
  681, L77

\bibitem[{{Narayanan} {et~al.}(2008{\natexlab{b}}){Narayanan}, {Cox},
  {Shirley}, {Dav{\'e}}, {Hernquist}, \& {Walker}}]{narayanan08b}
{Narayanan}, D., {Cox}, T.~J., {Shirley}, Y., {Dav{\'e}}, R., {Hernquist}, L.,
  \& {Walker}, C.~K. 2008{\natexlab{b}}, \apj, 684, 996

\bibitem[{{Narayanan} {et~al.}(2011{\natexlab{a}}){Narayanan}, {Krumholz},
  {Ostriker}, \& {Hernquist}}]{narayanan11b}
{Narayanan}, D., {Krumholz}, M., {Ostriker}, E.~C., \& {Hernquist}, L.
  2011{\natexlab{a}}, \mnras, submitted, arXiv:1110.3791

\bibitem[{{Narayanan} {et~al.}(2011{\natexlab{b}}){Narayanan}, {Krumholz},
  {Ostriker}, \& {Hernquist}}]{narayanan11a}
---. 2011{\natexlab{b}}, \mnras, in press, arXiv:1104.4118

\bibitem[{{Ostriker} {et~al.}(2010){Ostriker}, {McKee}, \&
  {Leroy}}]{ostriker10a}
{Ostriker}, E.~C., {McKee}, C.~F., \& {Leroy}, A.~K. 2010, \apj, 721, 975

\bibitem[{{Ostriker} \& {Shetty}(2011)}]{ostriker11a}
{Ostriker}, E.~C., \& {Shetty}, R. 2011, \apj, 731, 41

\bibitem[{{Padoan} \& {Nordlund}(2011)}]{padoan11a}
{Padoan}, P., \& {Nordlund}, {\AA}. 2011, \apj, 730, 40

\bibitem[{{Powell} {et~al.}(2011){Powell}, {Bournaud}, {Chapon}, {Devriendt},
  {Slyz}, \& {Teyssier}}]{powell11a}
{Powell}, L.~C., {Bournaud}, F., {Chapon}, D., {Devriendt}, J., {Slyz}, A., \&
  {Teyssier}, R. 2011, ArXiv e-prints

\bibitem[{{Robertson} \& {Kravtsov}(2008)}]{robertson08b}
{Robertson}, B.~E., \& {Kravtsov}, A.~V. 2008, \apj, 680, 1083

\bibitem[{{Roman-Duval} {et~al.}(2010){Roman-Duval}, {Jackson}, {Heyer},
  {Rathborne}, \& {Simon}}]{roman-duval10a}
{Roman-Duval}, J., {Jackson}, J.~M., {Heyer}, M., {Rathborne}, J., \& {Simon},
  R. 2010, \apj, 723, 492

\bibitem[{{Rosolowsky}(2005)}]{rosolowsky05b}
{Rosolowsky}, E. 2005, \pasp, 117, 1403

\bibitem[{{Saintonge} {et~al.}(2011{\natexlab{a}}){Saintonge}, {Kauffmann},
  {Kramer}, {Tacconi}, {Buchbender}, {Catinella}, {Fabello},
  {Graci{\'a}-Carpio}, {Wang}, {Cortese}, {Fu}, {Genzel}, {Giovanelli}, {Guo},
  {Haynes}, {Heckman}, {Krumholz}, {Lemonias}, {Li}, {Moran},
  {Rodriguez-Fernandez}, {Schiminovich}, {Schuster}, \&
  {Sievers}}]{saintonge11a}
{Saintonge}, A., {et~al.} 2011{\natexlab{a}}, \mnras, 960

\bibitem[{{Saintonge} {et~al.}(2011{\natexlab{b}}){Saintonge}, {Kauffmann},
  {Wang}, {Kramer}, {Tacconi}, {Buchbender}, {Catinella}, {Graci{\'a}-Carpio},
  {Cortese}, {Fabello}, {Fu}, {Genzel}, {Giovanelli}, {Guo}, {Haynes},
  {Heckman}, {Krumholz}, {Lemonias}, {Li}, {Moran}, {Rodriguez-Fernandez},
  {Schiminovich}, {Schuster}, \& {Sievers}}]{saintonge11b}
---. 2011{\natexlab{b}}, \mnras, 964

\bibitem[{{Saitoh} {et~al.}(2010){Saitoh}, {Daisaka}, {Kokubo}, {Makino},
  {Oakmoto}, {Tomisaka}, {Wada}, \& {Yoshida}}]{saitoh10a}
{Saitoh}, T.~R., {Daisaka}, H., {Kokubo}, E., {Makino}, J., {Oakmoto}, T.,
  {Tomisaka}, K., {Wada}, K., \& {Yoshida}, N. 2010, in Astronomical Society of
  the Pacific Conference Series, Vol. 423, Galaxy Wars: Stellar Populations and
  Star Formation in Interacting Galaxies, ed. {B.~Smith, J.~Higdon, S.~Higdon,
  \& N.~Bastian}, 185--+

\bibitem[{{Schenck} {et~al.}(2011){Schenck}, {Shirley}, {Reiter}, \&
  {Juneau}}]{schenck11a}
{Schenck}, D.~E., {Shirley}, Y.~L., {Reiter}, M., \& {Juneau}, S. 2011, \aj,
  142, 94

\bibitem[{{Schmidt}(1959)}]{schmidt59a}
{Schmidt}, M. 1959, \apj, 129, 243

\bibitem[{{Scoville} {et~al.}(1997){Scoville}, {Yun}, \&
  {Bryant}}]{scoville97a}
{Scoville}, N.~Z., {Yun}, M.~S., \& {Bryant}, P.~M. 1997, \apj, 484, 702

\bibitem[{{Shetty} \& {Ostriker}(2008)}]{shetty08a}
{Shetty}, R., \& {Ostriker}, E.~C. 2008, \apj, 684, 978

\bibitem[{{Silk}(1997)}]{silk97a}
{Silk}, J. 1997, \apj, 481, 703

\bibitem[{{Silk} \& {Norman}(2009)}]{silk09a}
{Silk}, J., \& {Norman}, C. 2009, \apj, 700, 262

\bibitem[{{Solomon} {et~al.}(1997){Solomon}, {Downes}, {Radford}, \&
  {Barrett}}]{solomon97a}
{Solomon}, P.~M., {Downes}, D., {Radford}, S.~J.~E., \& {Barrett}, J.~W. 1997,
  \apj, 478, 144

\bibitem[{{Solomon} {et~al.}(1987){Solomon}, {Rivolo}, {Barrett}, \&
  {Yahil}}]{solomon87a}
{Solomon}, P.~M., {Rivolo}, A.~R., {Barrett}, J., \& {Yahil}, A. 1987, \apj,
  319, 730

\bibitem[{{Springel} \& {Hernquist}(2003)}]{springel03a}
{Springel}, V., \& {Hernquist}, L. 2003, \mnras, 339, 312

\bibitem[{{Tacconi} {et~al.}(2010){Tacconi}, {Genzel}, {Neri}, {Cox}, {Cooper},
  {Shapiro}, {Bolatto}, {Bouch{\'e}}, {Bournaud}, {Burkert}, {Combes},
  {Comerford}, {Davis}, {Schreiber}, {Garcia-Burillo}, {Gracia-Carpio}, {Lutz},
  {Naab}, {Omont}, {Shapley}, {Sternberg}, \& {Weiner}}]{tacconi10a}
{Tacconi}, L.~J., {et~al.} 2010, \nat, 463, 781

\bibitem[{{Tacconi} {et~al.}(2008){Tacconi}, {Genzel}, {Smail}, {Neri},
  {Chapman}, {Ivison}, {Blain}, {Cox}, {Omont}, {Bertoldi}, {Greve},
  {F{\"o}rster Schreiber}, {Genel}, {Lutz}, {Swinbank}, {Shapley}, {Erb},
  {Cimatti}, {Daddi}, \& {Baker}}]{tacconi08a}
---. 2008, \apj, 680, 246

\bibitem[{{Tan}(2000)}]{tan00a}
{Tan}, J.~C. 2000, \apj, 536, 173

\bibitem[{{Teyssier} {et~al.}(2010){Teyssier}, {Chapon}, \&
  {Bournaud}}]{teyssier10a}
{Teyssier}, R., {Chapon}, D., \& {Bournaud}, F. 2010, \apjl, 720, L149

\bibitem[{{Thompson} {et~al.}(2005){Thompson}, {Quataert}, \&
  {Murray}}]{thompson05a}
{Thompson}, T.~A., {Quataert}, E., \& {Murray}, N. 2005, \apj, 630, 167

\bibitem[{{Walter} {et~al.}(2008){Walter}, {Brinks}, {de Blok}, {Bigiel},
  {Kennicutt}, {Thornley}, \& {Leroy}}]{walter08a}
{Walter}, F., {Brinks}, E., {de Blok}, W.~J.~G., {Bigiel}, F., {Kennicutt},
  R.~C., {Thornley}, M.~D., \& {Leroy}, A. 2008, \aj, 136, 2563

\bibitem[{{Wong} {et~al.}(2011){Wong}, {Hughes}, {Ott}, {Muller}, {Pineda},
  {Bernard}, {Chu}, {Fukui}, {Gruendl}, {Henkel}, {Kawamura}, {Klein},
  {Looney}, {Maddison}, {Mizuno}, {Paradis}, {Seale}, \& {Welty}}]{wong11a}
{Wong}, T., {et~al.} 2011, ArXiv e-prints

\bibitem[{{Wu} {et~al.}(2005){Wu}, {Evans}, {Gao}, {Solomon}, {Shirley}, \&
  {Vanden Bout}}]{wu05a}
{Wu}, J., {Evans}, N.~J., {Gao}, Y., {Solomon}, P.~M., {Shirley}, Y.~L., \&
  {Vanden Bout}, P.~A. 2005, \apjl, 635, L173

\bibitem[{{Wu} {et~al.}(2010){Wu}, {Evans}, {Shirley}, \& {Knez}}]{wu10a}
{Wu}, J., {Evans}, II, N.~J., {Shirley}, Y.~L., \& {Knez}, C. 2010, \apjs, 188,
  313

\bibitem[{{Young} {et~al.}(1989){Young}, {Xie}, {Kenney}, \& {Rice}}]{young89a}
{Young}, J.~S., {Xie}, S., {Kenney}, J.~D.~P., \& {Rice}, W.~L. 1989, \apjs,
  70, 699

\end{thebibliography}

\clearpage

\begin{appendix}

\section{Derived Quantities for Observed Data}

In the appendix we summarize the observed data and the quantities we derive from it for Galactic molecular clouds (Table \ref{tab:galdata}), unresolved local galaxies (Table \ref{tab:unresolved_local}), and unresolved high redshift galaxies (Table \ref{tab:unresolved_hi_z}).

\LongTables

\begin{deluxetable}{lrrrrrrrrr}
\tablecolumns{10}
\tablewidth{0pc}
\tablecaption{Galactic Data Set\label{tab:galdata}}
\tablehead{
\colhead{Object} &
\colhead{$M$} &
\colhead{$A$} &
\colhead{$\Sigma$} &
\colhead{$\dot{M}_*$} &
\colhead{$\dot{\Sigma}_*$} &
\colhead{$\rho/\mu_{\rm H}$\tablenotemark{a}} &
\colhead{$\tff$} &
\colhead{$\Sigma/\tff$\tablenotemark{b}} &
\colhead{$100\epsff$\tablenotemark{c}} \\
\colhead{} &
\colhead{($\msun$)} &
\colhead{(pc$^2$)} &
\colhead{($\msun/{\rm pc}^2$)} &
\colhead{($\msun/{\rm Myr}$)} &
\colhead{($\msun/{\rm pc}^2/{\rm Myr}$)} &
\colhead{($10^3$ cm$^{-3}$)} &
\colhead{(Myr)} &
\colhead{($\msun/{\rm pc}^2/{\rm Myr}$)} &
\colhead{} \\
}
\startdata
\cutinhead{Data from \citet{heiderman10a}} \\
Chameleon II &      637 &   9.91 &  64.3 &   6.0 & 0.61 & 0.78 & 1.55 &  41.4 &  1.46 \\
Lupus I &      513 &   8.86 &  57.9 &   3.2 & 0.37 & 0.75 & 1.59 &  36.4 &  1.01 \\
Lupus III &      912 &  15.40 &  59.2 &  17.0 & 1.10 & 0.58 & 1.81 &  32.7 &  3.37 \\
Lupus IV &      189 &   2.52 &  75.0 &   3.0 & 1.19 & 1.81 & 1.02 &  73.4 &  1.62 \\
Ophiuchus &     3120 &  29.60 & 105.0 &  72.5 & 2.45 & 0.74 & 1.60 &  65.7 &  3.71 \\
Perseus &     6590 &  73.20 &  90.0 &  96.2 & 1.31 & 0.40 & 2.17 &  41.6 &  3.16 \\
Serpens &     2340 &  17.00 & 138.0 &  56.0 & 3.29 & 1.28 & 1.21 & 113.7 &  2.91 \\
Auriga N &      224 &   2.41 &  92.9 &   0.5 & 0.21 & 2.29 & 0.91 & 102.3 &  0.20 \\
Auriga &     4620 &  50.00 &  92.4 &  42.7 & 0.85 & 0.50 & 1.94 &  47.6 &  1.80 \\
Cepheus &     2610 &  38.00 &  68.7 &  29.5 & 0.78 & 0.43 & 2.10 &  32.7 &  2.38 \\
Chameleon III &     1330 &  28.00 &  47.5 &   1.0 & 0.04 & 0.34 & 2.34 &  20.3 &  0.18 \\
Chameleon I &      857 &   9.41 &  91.1 &  22.2 & 2.36 & 1.14 & 1.29 &  70.7 &  3.34 \\
Corona Australis &      279 &   3.03 &  92.1 &  10.2 & 3.37 & 2.03 & 0.97 &  95.4 &  3.53 \\
IC5146E &     3370 &  61.40 &  54.9 &  23.2 & 0.38 & 0.27 & 2.65 &  20.7 &  1.83 \\
IC5146NW &     5180 &  87.60 &  59.1 &   9.5 & 0.11 & 0.24 & 2.79 &  21.1 &  0.51 \\
Lupus VI &      455 &   6.74 &  67.5 &  11.2 & 1.66 & 1.00 & 1.38 &  49.0 &  3.39 \\
Lupus V &      705 &  11.70 &  60.3 &  10.7 & 0.92 & 0.68 & 1.67 &  36.1 &  2.54 \\
Musca &      335 &   6.82 &  49.1 &   3.0 & 0.44 & 0.72 & 1.62 &  30.3 &  1.45 \\
Scorpius &      621 &   7.29 &  85.2 &   2.5 & 0.34 & 1.21 & 1.25 &  68.2 &  0.50 \\
Serpens-Aquila &    24400 & 179.00 & 136.0 & 360.0 & 2.01 & 0.39 & 2.20 &  61.7 &  3.25 \\
\cutinhead{Data from \citet{lada10a}} \\
Orion A, $A_K = 0.1$ mag\tablenotemark{d} &        67714 &  2335.0 &  29.0 & 715.0 & 0.31 & 0.023\tablenotemark{d} & 9.07\tablenotemark{d} &  3.2\tablenotemark{d} &  9.57\tablenotemark{d} \\
Orion B, $A_K = 0.1$ mag\tablenotemark{d} &        71828 &  2476.8 &  29.0 & 159.0 & 0.06 & 0.022\tablenotemark{d} & 9.20\tablenotemark{d} &  3.2\tablenotemark{d} &  2.04\tablenotemark{d} \\
California, $A_K = 0.1$ mag\tablenotemark{d} &        99930 &  3445.9 &  29.0 &  70.0 & 0.02 & 0.019\tablenotemark{d} & 9.99\tablenotemark{d} &  2.9\tablenotemark{d} &  0.70\tablenotemark{d} \\
Perseus, $A_K = 0.1$ mag\tablenotemark{d} &        18438 &   635.8 &  29.0 & 150.0 & 0.24 & 0.044\tablenotemark{d} & 6.55\tablenotemark{d} &  4.4\tablenotemark{d} &  5.33\tablenotemark{d} \\
Taurus, $A_K = 0.1$ mag\tablenotemark{d} &        14964 &   516.0 &  29.0 &  84.0 & 0.16 & 0.049\tablenotemark{d} & 6.22\tablenotemark{d} &  4.7\tablenotemark{d} &  3.49\tablenotemark{d} \\
Ophiuchus, $A_K = 0.1$ mag\tablenotemark{d} &        14165 &   488.4 &  29.0 &  79.0 & 0.16 & 0.050\tablenotemark{d} & 6.13\tablenotemark{d} &  4.7\tablenotemark{d} &  3.42\tablenotemark{d} \\
RCrA, $A_K = 0.1$ mag\tablenotemark{d} &         1137 &    39.2 &  29.0 &  25.0 & 0.64 & 0.177\tablenotemark{d} & 3.26\tablenotemark{d} &  8.9\tablenotemark{d} &  7.18\tablenotemark{d} \\
Pipe, $A_K = 0.1$ mag &         7937 &   273.7 &  29.0 &   5.0 & 0.02 & 0.067 & 5.30 &  5.5 &  0.33 \\
Lupus 3, $A_K = 0.1$ mag &         2157 &    74.4 &  29.0 &  17.0 & 0.23 & 0.129 & 3.83 &  7.6 &  3.02 \\
Lupus 4, $A_K = 0.1$ mag &         1379 &    47.6 &  29.0 &   3.0 & 0.06 & 0.161 & 3.42 &  8.5 &  0.74 \\
Lupus 1, $A_K = 0.1$ mag &          787 &    27.1 &  29.0 &   3.0 & 0.11 & 0.213 & 2.98 &  9.7 &  1.13 \\
Orion A, $A_K = 0.8$ mag &        13721 &  59.14 & 232.0 & 238.3 &   4.0 &  1.16 & 1.28 & 181.4 &  2.22 \\
Orion B, $A_K = 0.8$ mag &         7261 &  31.30 & 232.0 &  53.0 &   1.7 &  1.59 & 1.09 & 212.7 &  0.80 \\
California, $A_K = 0.8$ mag &         3199 &  13.79 & 232.0 &  23.3 &   1.7 &  2.39 & 0.89 & 261.1 &  0.65 \\
Perseus, $A_K = 0.8$ mag &         1880 &   8.10 & 232.0 &  50.0 &   6.2 &  3.12 & 0.78 & 298.2 &  2.07 \\
Taurus, $A_K = 0.8$ mag &         1766 &   7.61 & 232.0 &  28.0 &   3.7 &  3.22 & 0.77 & 302.9 &  1.21 \\
Ophiuchus, $A_K = 0.8$ mag &         1296 &   5.59 & 232.0 &  26.3 &   4.7 &  3.76 & 0.71 & 327.3 &  1.44 \\
RCrA, $A_K = 0.8$ mag &          258 &   1.11 & 232.0 &   8.3 &   7.5 &  8.43 & 0.47 & 490.0 &  1.53 \\
Pipe, $A_K = 0.8$ mag &          178 &   0.77 & 232.0 &   1.7 &   2.2 & 10.15 & 0.43 & 537.6 &  0.40 \\
Lupus 3, $A_K = 0.8$ mag &          163 &   0.70 & 232.0 &   5.7 &   8.1 & 10.61 & 0.42 & 549.6 &  1.47 \\
Lupus 4, $A_K = 0.8$ mag &          124 &   0.53 & 232.0 &   1.0 &   1.9 & 12.16 & 0.39 & 588.5 &  0.32 \\
Lupus 1, $A_K = 0.8$ mag &           75 &   0.32 & 232.0 &   1.0 &   3.1 & 15.64 & 0.35 & 667.3 &  0.46 \\
\enddata
\tablenotetext{a}{Computed by $\rho/m_{\rm H}=(3\sqrt{\pi}/4)M/A^{3/2}/m_{\rm H}$, where $\mu_{\rm H}=2.34\times 10^{-24}$ g is the mean mass per H nucleus for a gas of standard cosmic composition}
\tablenotetext{b}{Computed by $\tff=\sqrt{3\pi/32 G \rho}$}
\tablenotetext{c}{Computed by $\epsff=\dot{\Sigma}_*/(\Sigma/\tff)$}
\tablenotetext{d}{These clouds are known to be highly filamentary, so the values of $\rho$ are likely to be systematically underestimated, and the values of $\tff$ and $\epsff$ are likely to be systematically overestimated.}
\end{deluxetable}

\clearpage

\begin{deluxetable}{lrrrrrrrrr}
\tablecaption{Unresolved Local Extragalactic Data Set\label{tab:unresolved_local}}
\tablehead{
\colhead{Object} &
\colhead{D/SB\tablenotemark{a}} &
\colhead{$\log \Sigma$} &
\colhead{$\log \torb$} &
\colhead{$\log \Sigma/\torb$} &
\colhead{$\log \dot{\Sigma}_*$} &
\colhead{$\log \tffgmc$\tablenotemark{b}} &
\colhead{$\log \tfft$\tablenotemark{c}} &
\colhead{$\log \Sigma/\tff$\tablenotemark{d}} &
\colhead{$100\epsff$\tablenotemark{e}}
\\
\colhead{} & 
\colhead{} &
\colhead{($\msun/{\rm pc}^2$)} &
\colhead{(Myr)} &
\colhead{($\msun/{\rm pc}^2/{\rm Myr}$)} &
\colhead{($\msun/{\rm pc}^2/{\rm Myr}$)} &
\colhead{(Myr)} &
\colhead{(Myr)} &
\colhead{($\msun/{\rm pc}^2/{\rm Myr}$)} &
\colhead{} \\
}
\startdata
NGC 253 &  SB  & $2.60$ & $1.18$ & $ 1.42$ & $ 1.00$ & $ 1.63$ & $-0.28$ & $ 2.88$ & $ 1.32$ \\
NGC 520 &  SB  & $1.85$ & $1.66$ & $ 0.19$ & $-0.18$ & $ 1.82$ & $ 0.21$ & $ 1.64$ & $ 1.50$ \\
NGC 660 &  SB  & $2.91$ & $1.41$ & $ 1.50$ & $ 0.86$ & $ 1.55$ & $-0.04$ & $ 2.95$ & $ 0.81$ \\
NGC 828 &  SB  & $1.86$ & $1.34$ & $ 0.52$ & $-0.82$ & $ 1.82$ & $-0.11$ & $ 1.97$ & $ 0.16$ \\
NGC 891 &  SB  & $1.92$ & $1.46$ & $ 0.46$ & $-0.44$ & $ 1.80$ & $ 0.01$ & $ 1.91$ & $ 0.44$ \\
NGC 1097 &  SB  & $2.97$ & $1.28$ & $ 1.69$ & $ 1.55$ & $ 1.54$ & $-0.17$ & $ 3.14$ & $ 2.53$ \\
NGC 1614 &  SB  & $1.90$ & $1.72$ & $ 0.18$ & $-0.16$ & $ 1.81$ & $ 0.26$ & $ 1.64$ & $ 1.59$ \\
NGC 1808 &  SB  & $2.08$ & $1.51$ & $ 0.57$ & $ 0.60$ & $ 1.76$ & $ 0.05$ & $ 2.03$ & $ 3.71$ \\
NGC 2146 &  SB  & $2.77$ & $0.95$ & $ 1.82$ & $ 1.24$ & $ 1.59$ & $-0.50$ & $ 3.27$ & $ 0.93$ \\
NGC 2623 &  SB  & $2.38$ & $2.09$ & $ 0.29$ & $ 0.44$ & $ 1.69$ & $ 0.63$ & $ 1.75$ & $ 4.91$ \\
NGC 2903 &  SB  & $2.08$ & $1.11$ & $ 0.97$ & $-0.00$ & $ 1.76$ & $-0.34$ & $ 2.42$ & $ 0.38$ \\
NGC 3034 &  SB  & $2.15$ & $1.30$ & $ 0.85$ & $-0.13$ & $ 1.74$ & $-0.15$ & $ 2.30$ & $ 0.37$ \\
NGC 3079 &  SB  & $1.53$ & $1.58$ & $-0.05$ & $-1.01$ & $ 1.90$ & $ 0.13$ & $ 1.40$ & $ 0.38$ \\
NGC 3256 &  SB  & $1.50$ & $1.15$ & $ 0.35$ & $-0.42$ & $ 1.91$ & $-0.31$ & $ 1.81$ & $ 0.59$ \\
NGC 3351 &  SB  & $1.74$ & $1.64$ & $ 0.10$ & $-1.35$ & $ 1.85$ & $ 0.19$ & $ 1.55$ & $ 0.13$ \\
NGC 3504 &  SB  & $2.12$ & $1.32$ & $ 0.80$ & $ 0.06$ & $ 1.75$ & $-0.13$ & $ 2.25$ & $ 0.64$ \\
NGC 3627 &  SB  & $3.36$ & $1.38$ & $ 1.98$ & $ 1.63$ & $ 1.44$ & $-0.07$ & $ 3.43$ & $ 1.57$ \\
NGC 3690 &  SB  & $1.51$ & $1.97$ & $-0.46$ & $-0.54$ & $ 1.90$ & $ 0.52$ & $ 0.99$ & $ 2.93$ \\
NGC 4736 &  SB  & $3.20$ & $1.36$ & $ 1.84$ & $ 1.30$ & $ 1.48$ & $-0.09$ & $ 3.29$ & $ 1.01$ \\
NGC 5194 &  SB  & $3.06$ & $1.38$ & $ 1.68$ & $ 1.43$ & $ 1.52$ & $-0.07$ & $ 3.13$ & $ 1.97$ \\
NGC 5236 &  SB  & $1.71$ & $1.89$ & $-0.18$ & $-0.51$ & $ 1.85$ & $ 0.44$ & $ 1.27$ & $ 1.65$ \\
NGC 6240 &  SB  & $1.98$ & $2.13$ & $-0.15$ & $ 0.08$ & $ 1.79$ & $ 0.67$ & $ 1.31$ & $ 5.91$ \\
NGC 6946 &  SB  & $4.01$ & $0.78$ & $ 3.23$ & $ 2.74$ & $ 1.28$ & $-0.67$ & $ 4.68$ & $ 1.13$ \\
NGC 7252 &  SB  & $1.66$ & $2.23$ & $-0.57$ & $ 0.24$ & $ 1.87$ & $ 0.77$ & $ 0.89$ & $22.37$ \\
NGC 7552 &  SB  & $4.25$ & $1.91$ & $ 2.34$ & $ 2.87$ & $ 1.22$ & $ 0.46$ & $ 3.79$ & $11.97$ \\
IC 342 &  SB  & $3.15$ & $1.28$ & $ 1.87$ & $ 1.53$ & $ 1.49$ & $-0.17$ & $ 3.32$ & $ 1.60$ \\
NGC 224 &  D  & $0.68$ & $2.66$ & $-1.98$ & $-3.37$ & $ 1.41$ & $ 1.36$ & $-0.68$ & $ 0.20$ \\
NGC 598 &  D  & $1.03$ & $2.60$ & $-1.57$ & $-2.71$ & $ 1.33$ & $ 1.30$ & $-0.27$ & $ 0.36$ \\
NGC 628 &  D  & $0.94$ & $2.90$ & $-1.96$ & $-3.08$ & $ 1.35$ & $ 1.60$ & $-0.41$ & $ 0.21$ \\
NGC 772 &  D  & $0.91$ & $2.86$ & $-1.95$ & $-2.68$ & $ 1.36$ & $ 1.56$ & $-0.45$ & $ 0.58$ \\
NGC 925 &  D  & $1.33$ & $2.30$ & $-0.97$ & $-1.04$ & $ 1.25$ & $ 1.00$ & $ 0.33$ & $ 4.23$ \\
NGC 1058 &  D  & $0.91$ & $2.87$ & $-1.96$ & $-2.16$ & $ 1.36$ & $ 1.57$ & $-0.45$ & $ 1.92$ \\
NGC 1569 &  D  & $0.88$ & $2.54$ & $-1.66$ & $-2.39$ & $ 1.36$ & $ 1.24$ & $-0.36$ & $ 0.93$ \\
NGC 2336 &  D  & $0.97$ & $2.20$ & $-1.23$ & $-3.23$ & $ 1.34$ & $ 0.90$ & $ 0.07$ & $ 0.05$ \\
NGC 2403 &  D  & $0.86$ & $2.49$ & $-1.63$ & $-2.55$ & $ 1.37$ & $ 1.19$ & $-0.33$ & $ 0.60$ \\
NGC 2841 &  D  & $0.98$ & $2.11$ & $-1.13$ & $-1.90$ & $ 1.34$ & $ 0.81$ & $ 0.17$ & $ 0.85$ \\
NGC 2903 &  D  & $0.85$ & $2.43$ & $-1.58$ & $-2.74$ & $ 1.37$ & $ 1.13$ & $-0.28$ & $ 0.34$ \\
NGC 2976 &  D  & $1.14$ & $2.40$ & $-1.26$ & $-1.38$ & $ 1.30$ & $ 1.10$ & $ 0.04$ & $ 3.74$ \\
NGC 3031 &  D  & $0.81$ & $2.60$ & $-1.79$ & $-2.80$ & $ 1.38$ & $ 1.30$ & $-0.49$ & $ 0.49$ \\
NGC 3310 &  D  & $0.93$ & $2.43$ & $-1.50$ & $-2.79$ & $ 1.35$ & $ 1.13$ & $-0.20$ & $ 0.25$ \\
NGC 3338 &  D  & $0.88$ & $2.51$ & $-1.63$ & $-2.70$ & $ 1.36$ & $ 1.20$ & $-0.32$ & $ 0.42$ \\
NGC 3368 &  D  & $1.22$ & $2.52$ & $-1.30$ & $-2.15$ & $ 1.28$ & $ 1.22$ & $ 0.00$ & $ 0.70$ \\
NGC 3486 &  D  & $1.16$ & $2.69$ & $-1.53$ & $-1.97$ & $ 1.29$ & $ 1.39$ & $-0.13$ & $ 1.45$ \\
NGC 3521 &  D  & $0.99$ & $2.34$ & $-1.35$ & $-2.25$ & $ 1.34$ & $ 1.04$ & $-0.05$ & $ 0.63$ \\
NGC 3631 &  D  & $1.06$ & $2.51$ & $-1.45$ & $-2.52$ & $ 1.32$ & $ 1.20$ & $-0.14$ & $ 0.42$ \\
NGC 3675 &  D  & $1.06$ & $2.48$ & $-1.42$ & $-2.20$ & $ 1.32$ & $ 1.17$ & $-0.11$ & $ 0.82$ \\
NGC 3726 &  D  & $1.13$ & $2.59$ & $-1.46$ & $-2.51$ & $ 1.30$ & $ 1.29$ & $-0.16$ & $ 0.44$ \\
NGC 3893 &  D  & $1.39$ & $2.54$ & $-1.15$ & $-1.94$ & $ 1.24$ & $ 1.24$ & $ 0.15$ & $ 0.80$ \\
NGC 3938 &  D  & $0.59$ & $2.68$ & $-2.09$ & $-2.60$ & $ 1.44$ & $ 1.38$ & $-0.79$ & $ 1.53$ \\
NGC 4178 &  D  & $1.02$ & $2.48$ & $-1.46$ & $-2.11$ & $ 1.33$ & $ 1.17$ & $-0.15$ & $ 1.10$ \\
NGC 4189 &  D  & $1.21$ & $2.68$ & $-1.47$ & $-1.98$ & $ 1.28$ & $ 1.38$ & $-0.07$ & $ 1.22$ \\
NGC 4254 &  D  & $1.14$ & $2.65$ & $-1.51$ & $-2.31$ & $ 1.30$ & $ 1.35$ & $-0.16$ & $ 0.70$ \\
NGC 4258 &  D  & $0.63$ & $2.51$ & $-1.88$ & $-3.12$ & $ 1.43$ & $ 1.20$ & $-0.57$ & $ 0.28$ \\
NGC 4294 &  D  & $1.08$ & $2.62$ & $-1.54$ & $-3.04$ & $ 1.31$ & $ 1.32$ & $-0.23$ & $ 0.15$ \\
NGC 4299 &  D  & $1.09$ & $2.52$ & $-1.43$ & $-2.45$ & $ 1.31$ & $ 1.22$ & $-0.13$ & $ 0.47$ \\
NGC 4303 &  D  & $0.99$ & $2.46$ & $-1.47$ & $-2.22$ & $ 1.34$ & $ 1.16$ & $-0.17$ & $ 0.89$ \\
NGC 4321 &  D  & $1.01$ & $2.72$ & $-1.71$ & $-2.62$ & $ 1.33$ & $ 1.41$ & $-0.32$ & $ 0.50$ \\
NGC 4394 &  D  & $0.69$ & $2.53$ & $-1.84$ & $-2.76$ & $ 1.41$ & $ 1.23$ & $-0.54$ & $ 0.60$ \\
NGC 4402 &  D  & $1.52$ & $2.20$ & $-0.68$ & $-2.17$ & $ 1.20$ & $ 0.90$ & $ 0.62$ & $ 0.16$ \\
NGC 4501 &  D  & $0.61$ & $2.70$ & $-2.09$ & $-3.02$ & $ 1.43$ & $ 1.40$ & $-0.79$ & $ 0.58$ \\
NGC 4519 &  D  & $0.83$ & $2.67$ & $-1.84$ & $-2.80$ & $ 1.38$ & $ 1.37$ & $-0.54$ & $ 0.55$ \\
NGC 4535 &  D  & $0.81$ & $2.45$ & $-1.64$ & $-2.56$ & $ 1.38$ & $ 1.14$ & $-0.33$ & $ 0.59$ \\
NGC 4548 &  D  & $0.73$ & $2.34$ & $-1.61$ & $-2.35$ & $ 1.40$ & $ 1.04$ & $-0.31$ & $ 0.91$ \\
NGC 4561 &  D  & $1.04$ & $2.54$ & $-1.50$ & $-2.46$ & $ 1.32$ & $ 1.24$ & $-0.20$ & $ 0.55$ \\
NGC 4569 &  D  & $1.06$ & $2.43$ & $-1.37$ & $-2.22$ & $ 1.32$ & $ 1.13$ & $-0.07$ & $ 0.70$ \\
NGC 4571 &  D  & $1.10$ & $2.54$ & $-1.44$ & $-2.30$ & $ 1.31$ & $ 1.24$ & $-0.14$ & $ 0.69$ \\
NGC 4579 &  D  & $0.94$ & $2.51$ & $-1.57$ & $-2.62$ & $ 1.35$ & $ 1.20$ & $-0.26$ & $ 0.44$ \\
NGC 4639 &  D  & $0.25$ & $2.40$ & $-2.15$ & $-3.79$ & $ 1.52$ & $ 1.10$ & $-0.85$ & $ 0.11$ \\
NGC 4647 &  D  & $1.04$ & $2.51$ & $-1.47$ & $-1.77$ & $ 1.32$ & $ 1.20$ & $-0.16$ & $ 2.45$ \\
NGC 4651 &  D  & $0.65$ & $2.43$ & $-1.78$ & $-2.46$ & $ 1.42$ & $ 1.13$ & $-0.48$ & $ 1.04$ \\
NGC 4654 &  D  & $0.93$ & $2.89$ & $-1.96$ & $-2.88$ & $ 1.35$ & $ 1.58$ & $-0.42$ & $ 0.34$ \\
NGC 4689 &  D  & $1.17$ & $2.58$ & $-1.41$ & $-2.56$ & $ 1.29$ & $ 1.28$ & $-0.11$ & $ 0.35$ \\
NGC 4698 &  D  & $1.47$ & $2.53$ & $-1.06$ & $-2.02$ & $ 1.22$ & $ 1.23$ & $ 0.25$ & $ 0.53$ \\
NGC 4713 &  D  & $1.70$ & $2.45$ & $-0.75$ & $-1.65$ & $ 1.16$ & $ 1.14$ & $ 0.56$ & $ 0.62$ \\
NGC 4736 &  D  & $1.09$ & $2.94$ & $-1.85$ & $-2.70$ & $ 1.31$ & $ 1.64$ & $-0.22$ & $ 0.33$ \\
NGC 4826 &  D  & $1.03$ & $2.41$ & $-1.38$ & $-1.94$ & $ 1.33$ & $ 1.11$ & $-0.08$ & $ 1.38$ \\
NGC 5033 &  D  & $1.29$ & $2.46$ & $-1.17$ & $-2.15$ & $ 1.26$ & $ 1.16$ & $ 0.13$ & $ 0.52$ \\
NGC 5055 &  D  & $0.89$ & $2.23$ & $-1.34$ & $-2.32$ & $ 1.36$ & $ 0.93$ & $-0.04$ & $ 0.52$ \\
NGC 5194 &  D  & $1.11$ & $2.57$ & $-1.46$ & $-2.05$ & $ 1.31$ & $ 1.27$ & $-0.16$ & $ 1.27$ \\
NGC 5236 &  D  & $1.30$ & $2.54$ & $-1.24$ & $-2.12$ & $ 1.26$ & $ 1.24$ & $ 0.06$ & $ 0.66$ \\
NGC 5457 &  D  & $1.08$ & $2.76$ & $-1.68$ & $-2.57$ & $ 1.31$ & $ 1.46$ & $-0.23$ & $ 0.46$ \\
\enddata
\tablecomments{All data are taken from \citet{kennicutt98a}, adjusted to the same IMF and CO $X$ factor as the high-$z$ data following \citet{daddi10a}.}
\tablenotetext{a}{D=disk, SB = starburst}
\tablenotetext{b}{Computed from equation (\ref{eq:tffexgal1}) using $\sigma=8$ km s$^{-1}$ for disks and $\sigma=50$ km s$^{-1}$ for starbursts}
\tablenotetext{c}{Computed from equation (\ref{eq:tffexgal2}) using $Q=1$, $\beta=0$ for disks and $Q=1$, $\beta=0$ for starbursts}
\tablenotetext{d}{Computed using $\tff=\min(\tffgmc,\tfft)$}
\tablenotetext{e}{Computed from $\epsff = \dot{\Sigma}_* / (\Sigma/\min(\tffgmc,\tfft)$}
\end{deluxetable}

\clearpage

\begin{deluxetable}{lrrrrrrrrr}
\tablecolumns{10}
\tablewidth{0pc}
\tablecaption{Unresolved High-$z$ Extragalactic Data Set\label{tab:unresolved_hi_z}}
\tablehead{
\colhead{Object\tablenotemark{a}} &
\colhead{D/SB\tablenotemark{b}} &
\colhead{$\log \Sigma$} &
\colhead{$\log \torb$} &
\colhead{$\log \Sigma/\torb$} &
\colhead{$\log \dot{\Sigma}_*$} &
\colhead{$\log \tffgmc$\tablenotemark{c}} &
\colhead{$\log \tfft$\tablenotemark{d}} &
\colhead{$\log \Sigma/\tff$\tablenotemark{e}} &
\colhead{$100\epsff$\tablenotemark{f}}
\\
\colhead{} & 
\colhead{} &
\colhead{($\msun/{\rm pc}^2$)} &
\colhead{(Myr)} &
\colhead{($\msun/{\rm pc}^2/{\rm Myr}$)} &
\colhead{($\msun/{\rm pc}^2/{\rm Myr}$)} &
\colhead{(Myr)} &
\colhead{(Myr)} &
\colhead{($\msun/{\rm pc}^2/{\rm Myr}$)} &
\colhead{} \\
}
\startdata
\cutinhead{Data from \citet{genzel10a}} \\
Q2343-MD59 & D & $2.74$ & $2.34$ & $ 0.40$ & $-0.68$ & $ 1.60$ & $ 1.04$ & $ 1.71$ & $ 0.41$ \\
SMMJ02399-0136 & SB & $2.63$ & $1.73$ & $ 0.90$ & $ 0.84$ & $ 1.62$ & $ 0.28$ & $ 2.36$ & $ 3.04$ \\
SMMJ09431+4700 & SB & $3.53$ & $1.47$ & $ 2.06$ & $ 1.83$ & $ 1.40$ & $ 0.02$ & $ 3.52$ & $ 2.05$ \\
SMMJ105141+5719 & SB & $2.66$ & $1.62$ & $ 1.04$ & $ 1.04$ & $ 1.62$ & $ 0.17$ & $ 2.50$ & $ 3.49$ \\
SMMJ123549+6215 & SB & $4.00$ & $1.12$ & $ 2.88$ & $ 2.20$ & $ 1.28$ & $-0.33$ & $ 4.34$ & $ 0.73$ \\
SMMJ123634+6212 & SB & $2.58$ & $1.87$ & $ 0.71$ & $ 0.63$ & $ 1.64$ & $ 0.42$ & $ 2.17$ & $ 2.90$ \\
SMMJ123707+6214 & SB & $2.73$ & $1.74$ & $ 0.99$ & $ 1.01$ & $ 1.60$ & $ 0.29$ & $ 2.45$ & $ 3.65$ \\
SMMJ131201+4242 & SB & $2.99$ & $1.63$ & $ 1.36$ & $ 1.09$ & $ 1.53$ & $ 0.18$ & $ 2.82$ & $ 1.87$ \\
SMMJ131232+4239 & SB & $3.18$ & $1.56$ & $ 1.62$ & $ 1.29$ & $ 1.49$ & $ 0.11$ & $ 3.08$ & $ 1.63$ \\
SMMJ163650+4057 & SB & $3.42$ & $1.46$ & $ 1.96$ & $ 1.37$ & $ 1.43$ & $ 0.01$ & $ 3.42$ & $ 0.90$ \\
\cutinhead{Data from \citet{bouche07a}} \\
 \nodata &  SB  & $2.90$ & $1.80$ & $ 1.10$ & $ 1.10$ & $ 1.56$ & $ 0.35$ & $ 2.55$ & $ 3.51$ \\
 \nodata &  SB  & $3.10$ & $1.65$ & $ 1.45$ & $ 1.20$ & $ 1.51$ & $ 0.20$ & $ 2.90$ & $ 1.97$ \\
 \nodata &  SB  & $3.25$ & $1.80$ & $ 1.45$ & $ 1.15$ & $ 1.47$ & $ 0.35$ & $ 2.90$ & $ 1.76$ \\
 \nodata &  SB  & $3.45$ & $1.45$ & $ 2.00$ & $ 1.10$ & $ 1.42$ & $-0.00$ & $ 3.45$ & $ 0.44$ \\
 \nodata &  SB  & $2.30$ & $1.55$ & $ 0.75$ & $ 1.40$ & $ 1.71$ & $ 0.10$ & $ 2.20$ & $15.69$ \\
 \nodata &  SB  & $3.45$ & $1.55$ & $ 1.90$ & $ 1.70$ & $ 1.42$ & $ 0.10$ & $ 3.35$ & $ 2.22$ \\
 \nodata &  SB  & $3.30$ & $1.50$ & $ 1.80$ & $ 1.80$ & $ 1.46$ & $ 0.05$ & $ 3.25$ & $ 3.51$ \\
 \nodata &  SB  & $3.65$ & $1.40$ & $ 2.25$ & $ 1.80$ & $ 1.37$ & $-0.05$ & $ 3.70$ & $ 1.25$ \\
 \nodata &  SB  & $3.35$ & $1.25$ & $ 2.10$ & $ 2.00$ & $ 1.44$ & $-0.20$ & $ 3.55$ & $ 2.79$ \\
 \nodata &  SB  & $3.70$ & $1.45$ & $ 2.25$ & $ 1.95$ & $ 1.36$ & $-0.00$ & $ 3.70$ & $ 1.76$ \\
 \nodata &  SB  & $3.20$ & $1.55$ & $ 1.65$ & $ 2.10$ & $ 1.48$ & $ 0.10$ & $ 3.10$ & $ 9.90$ \\
 \nodata &  SB  & $3.70$ & $1.05$ & $ 2.65$ & $ 2.60$ & $ 1.36$ & $-0.40$ & $ 4.10$ & $ 3.13$ \\
 \nodata &  SB  & $4.00$ & $0.95$ & $ 3.05$ & $ 2.70$ & $ 1.28$ & $-0.50$ & $ 4.50$ & $ 1.57$ \\
\cutinhead{Data from \citet{daddi10a}} \\
 \nodata &  D  & $2.39$ & $2.08$ & $ 0.31$ & $-0.47$ & $ 1.68$ & $ 0.78$ & $ 1.61$ & $ 0.82$ \\
 \nodata &  D  & $2.53$ & $2.08$ & $ 0.44$ & $-0.36$ & $ 1.65$ & $ 0.78$ & $ 1.75$ & $ 0.78$ \\
 \nodata &  D  & $1.75$ & $2.08$ & $-0.33$ & $-1.27$ & $ 1.84$ & $ 0.78$ & $ 0.97$ & $ 0.57$ \\
\cutinhead{Data from \citet{daddi10b}} \\
BzK-4171 &  D  & $2.95$ & $1.84$ & $ 1.11$ & $ 0.08$ & $ 1.54$ & $ 0.54$ & $ 2.41$ & $ 0.46$ \\
BzK-21000 &  D  & $2.96$ & $1.62$ & $ 1.34$ & $ 0.19$ & $ 1.54$ & $ 0.32$ & $ 2.64$ & $ 0.36$ \\
BzK-16000 &  D  & $2.56$ & $1.83$ & $ 0.73$ & $-0.03$ & $ 1.64$ & $ 0.53$ & $ 2.03$ & $ 0.88$ \\
BzK-17999 &  D  & $2.66$ & $2.02$ & $ 0.64$ & $ 0.03$ & $ 1.62$ & $ 0.72$ & $ 1.94$ & $ 1.22$ \\
BzK-12591 &  D  & $2.53$ & $2.15$ & $ 0.38$ & $-0.04$ & $ 1.65$ & $ 0.84$ & $ 1.68$ & $ 1.89$ \\
BzK-25536 &  D  & $2.88$ & $2.25$ & $ 0.63$ & $ 0.05$ & $ 1.56$ & $ 0.95$ & $ 1.93$ & $ 1.31$ \\
\cutinhead{Data from \citet{tacconi10a}} \\
EGS13004291 &  D  & $2.85$ & $2.02$ & $ 0.83$ & $-0.28$ & $ 1.57$ & $ 0.72$ & $ 2.13$ & $ 0.39$ \\
EGS12007881 &  D  & $2.15$ & $2.33$ & $-0.17$ & $-0.72$ & $ 1.74$ & $ 1.02$ & $ 1.13$ & $ 1.42$ \\
EGS13017614 &  D  & $2.46$ & $2.03$ & $ 0.43$ & $-0.57$ & $ 1.67$ & $ 0.73$ & $ 1.73$ & $ 0.50$ \\
EGS13035123 &  D  & $2.30$ & $2.28$ & $ 0.01$ & $-0.61$ & $ 1.71$ & $ 0.98$ & $ 1.31$ & $ 1.20$ \\
EGS13004661 &  D  & $1.87$ & $2.21$ & $-0.34$ & $-0.52$ & $ 1.82$ & $ 0.90$ & $ 0.96$ & $ 3.26$ \\
EGS13003805 &  D  & $2.80$ & $2.12$ & $ 0.69$ & $-0.25$ & $ 1.58$ & $ 0.82$ & $ 1.99$ & $ 0.58$ \\
EGS12011767 &  D  & $1.83$ & $2.58$ & $-0.75$ & $-0.82$ & $ 1.82$ & $ 1.28$ & $ 0.55$ & $ 4.30$ \\
EGS12012083 &  D  & $2.44$ & $2.33$ & $ 0.11$ & $-0.17$ & $ 1.67$ & $ 1.03$ & $ 1.41$ & $ 2.62$ \\
EGS13011439 &  D  & $2.07$ & $2.12$ & $-0.05$ & $-0.62$ & $ 1.76$ & $ 0.82$ & $ 1.25$ & $ 1.35$ \\
HDF-BX1439 &  D  & $3.42$ & $1.67$ & $ 1.75$ & $ 0.41$ & $ 1.43$ & $ 0.36$ & $ 3.06$ & $ 0.23$ \\
Q1623-BX599 &  D  & $2.19$ & $2.28$ & $-0.09$ & $-0.60$ & $ 1.73$ & $ 0.98$ & $ 1.22$ & $ 1.55$ \\
Q1623-BX663 &  D  & $2.87$ & $2.29$ & $ 0.59$ & $-0.18$ & $ 1.56$ & $ 0.99$ & $ 1.89$ & $ 0.86$ \\
Q1700-MD69 &  D  & $3.20$ & $1.82$ & $ 1.38$ & $ 0.16$ & $ 1.48$ & $ 0.52$ & $ 2.69$ & $ 0.30$ \\
Q1700-MD94 &  D  & $2.01$ & $2.09$ & $-0.08$ & $-0.66$ & $ 1.78$ & $ 0.79$ & $ 1.22$ & $ 1.32$ \\
Q1700-MD174 &  D  & $2.45$ & $2.09$ & $ 0.36$ & $-0.49$ & $ 1.67$ & $ 0.79$ & $ 1.66$ & $ 0.71$ \\
Q1700-BX691 &  D  & $3.15$ & $1.79$ & $ 1.36$ & $ 0.03$ & $ 1.49$ & $ 0.49$ & $ 2.66$ & $ 0.23$ \\
\enddata
\tablenotetext{a}{A blank entry indicates the object is not identified by name in the source reference.}
\tablenotetext{b}{D=disk, SB = starburst}
\tablenotetext{c}{Computed from equation (\ref{eq:tffexgal1}) using $\sigma=50$ km s$^{-1}$}
\tablenotetext{d}{Computed from equation (\ref{eq:tffexgal2}) using $Q=1$, $\beta=0$ for disks and $Q=1$, $\beta=0$ for starbursts}
\tablenotetext{e}{Computed using $\tff=\min(\tffgmc,\tfft)$}
\tablenotetext{f}{Computed from $\epsff = \dot{\Sigma}_* / (\Sigma/\min[\tffgmc,\tfft])$}
\end{deluxetable}

\end{appendix}

\end{document}